\pdfoutput=1
\documentclass[letterpaper,journal]{IEEEtran}
\usepackage{amsmath,amsfonts,amssymb}
\usepackage{algorithmic}
\usepackage{algorithm}
\usepackage{array}
\usepackage[caption=false,font=footnotesize,labelfont=rm,textfont=rm,captionskip=0pt]{subfig}
\usepackage{textcomp}
\usepackage{stfloats}
\usepackage{url}
\usepackage[hidelinks]{hyperref}
\usepackage{graphicx}
\usepackage{cite}
\usepackage{booktabs}
\usepackage{tabularx}
\usepackage{tikz}
\usetikzlibrary{arrows.meta,positioning,shapes.geometric,calc,fit,backgrounds}

\usepackage{enumitem}
\setlist{noitemsep, topsep=2pt, parsep=0pt, partopsep=0pt}

\usepackage{etoolbox}
\AtBeginDocument{
  \setlength{\abovedisplayskip}{4pt plus 1pt minus 1pt}
  \setlength{\belowdisplayskip}{4pt plus 1pt minus 1pt}
  \setlength{\abovedisplayshortskip}{2pt plus 1pt minus 1pt}
  \setlength{\belowdisplayshortskip}{2pt plus 1pt minus 1pt}
}

\newcommand{\secref}[1]{Section~\ref{#1}}
\newcommand{\figref}[1]{Fig.~\ref{#1}}
\newcommand{\Figureref}[1]{Figure~\ref{#1}}
\newcommand{\eqnref}[1]{(\ref{#1})}
\newcommand{\tabref}[1]{Table~\ref{#1}}

\DeclareMathOperator*{\argmin}{argmin}
\begin{document}

\title{MiDiff: MIDI-Conditioned Diffusion Models\\for Drum Audio Enhancement}

\author{Elad~Aspis
        and~Sharon~Gannot,~\IEEEmembership{Fellow,~IEEE}%
\thanks{Elad Aspis and Sharon Gannot are with the Faculty of Engineering, Bar-Ilan University, Ramat-Gan 5290002, Israel (e-mail: elad.aspis@biu.ac.il, sharon.gannot@biu.ac.il).}%
\thanks{This work has been submitted to the IEEE for possible publication. Copyright may be transferred without notice, after which this version may no longer be accessible.}}

\markboth{IEEE/ACM Transactions on Audio, Speech, and Language Processing}%
{Aspis and Gannot: MiDiff: MIDI-Conditioned Diffusion Models for Drum Audio Enhancement}

\maketitle

\begin{abstract}
Bridging the semantic gap between perfect symbolic musical structure and severely degraded acoustic signals offers a powerful paradigm for high-fidelity audio restoration. In this work, we introduce MiDiff, a diffusion-based generative model for drum audio enhancement that conditions the reverse process on temporally aligned Musical Instrument Digital Interface (MIDI) piano rolls. We extend the Score-Based Generative Models for Speech Enhancement (SGMSE) framework in the complex Short-Time Fourier Transform (STFT) domain by integrating a pretrained MIDI encoder, multi-scale temporal alignment, and three complementary conditioning strategies: early injection, Feature-wise Linear Modulation (FiLM), and cross-attention. We further show that providing the same piano roll only through classifier guidance, driven by an Automatic Drum Transcription (ADT) model on the audio-only score, yields no measurable gain: the symbolic prior is effective when it conditions the score network. To relax the requirement for ground-truth MIDI at inference, we further propose a hybrid reverse schedule that transcribes the diffusion state online with the same noise-aware ADT network and hands over from the audio-only score to MiDiff at a tunable switch time, cutting test FAD by 47\% relative to the audio-only baseline without any symbolic input (switch time selected on the test set).
We train our model on the StemGMD dataset using a non-additive degradation pipeline that simulates room reverberation with echo, electret capsule frequency response, and soft-clipping saturation with noise.
MiDiff substantially outperforms the SGMSE baseline in terms of Fr\'echet Audio Distance (FAD) on both the validation and held-out test sets.
Ablation studies demonstrate that combining all three conditioning strategies yields superior performance and that authentic MIDI velocity dynamics are critical for effective symbolic conditioning.
Our method shows the strong potential of leveraging symbolic priors to reconstruct high-fidelity percussive audio.
Qualitative evaluation, audio demonstrations, and source code are available at \url{https://midiff-paper.github.io/midiff-project-page/}.
\end{abstract}


\section{Introduction}\label{sec:introduction}

\IEEEPARstart{T}{he} quality of drum recordings plays a vital role in how modern musical productions are perceived and how much impact they have. Achieving a professional-grade drum sound typically requires specialized acoustic environments, multi-microphone setups, high-end preamplifiers, and experienced audio engineers---resources that independent musicians and home-studio producers often cannot access. Unlike most acoustic instruments, which act as a single point source, a drum kit is a spatially distributed composite of distinct resonant bodies, each with its own extreme dynamic range and spectral footprint. Capturing this complexity creates practical challenges, including microphone bleed between closely spaced capsules~\cite{clifford2011microphone}, phase-coherence problems from different source-microphone distances, sympathetic resonances from poorly tuned drumheads, and excessive room reflections that wash out transient detail in untreated spaces.

Faced with these barriers, many drummers capture their performances using a single consumer device---typically a smartphone or action camera---resulting in mono or stereo tracks compromised by room reverberation, clipped transients, electret microphone coloration, and lossy codec compression. While electronic music production often uses pristine, pre-rendered sample libraries, restoring \emph{human-recorded} acoustic drum performances presents a distinct, challenging problem: live drum performances capture subtle expressive nuances, micro-timing variations, and organic velocity dynamics that static samples cannot easily replicate.

Aligned MIDI and a degraded acoustic capture of the same take are not hypothetical: trigger-equipped acoustic kits and DAW drum-replacement workflows both retain a MIDI stream recorded together with the acoustic signal rather than derived from it. This is the setting MiDiff targets.

Traditional audio enhancement methods primarily rely on discriminative deep learning to minimize element-wise error metrics, such as Mean Squared Error (MSE), in the time-frequency domain~\cite{xu2014regression}. However, regression-based objectives suffer from the well-documented ``regression to the mean'' phenomenon~\cite{pascual2017segan}, which causes transient smearing, phase inconsistency, and a loss of high-frequency detail. These flaws are particularly detrimental when processing percussive instruments. Generative diffusion models~\cite{ho2020denoising, kong2020diffwave, chen2020wavegrad} resolve these limitations by formulating enhancement as a stochastic reverse-time process that samples from a clean signal prior.

Despite recent advances in score-based diffusion models for audio enhancement~\cite{sgmse, song2021scorebased}, existing frameworks typically rely on degraded acoustic observations or supplementary visual cues~\cite{richter2023audio} for conditioning. When acoustic signals are severely degraded, pure audio-based conditioning provides ambiguous structural guidance, leading to spectral hallucinations or incomplete noise suppression. Fortunately, modern music production workflows routinely generate auxiliary symbolic data alongside raw audio. The Musical Instrument Digital Interface (MIDI)~\cite{midi1983specification} captures explicit temporal onset events, instrument identity, and note velocity dynamics. Because MIDI is recorded as a separate symbolic stream, obtained from drum triggers or electronic pads, sequencer tracks, or DAW programming, it is unaffected by acoustic degradation and provides a noise-free structural prior.

In this paper, we propose \textit{MiDiff}, a novel MIDI-conditioned diffusion framework for percussive audio enhancement. Our central hypothesis is that incorporating temporally aligned MIDI as a conditioning signal into a diffusion-based enhancement framework will improve the quality of restored drum audio, particularly with respect to transient preservation and instrument identity, compared to audio-only baselines. This hypothesis rests on the observation that MIDI already encodes \emph{when} each hit occurs and \emph{what} instrument was played, so the network can focus its capacity on recovering \emph{how} each hit sounds. Our key contributions are:

\begin{enumerate}
    \item \textbf{MIDI-Conditioned Score Network.} We extend the SGMSE framework~\cite{sgmse} to accept temporally aligned MIDI piano rolls as an additional conditioning input via a pretrained convolutional encoder and multi-scale temporal alignment modules.
    
    \item \textbf{Multi-Strategy Conditioning.} We investigate and compare three MIDI conditioning strategies: 1) early injection, 2) Feature-wise Linear Modulation (FiLM)~\cite{perez2018film}, and 3) cross-attention~\cite{vaswani2017attention}. We show that their combination yields superior enhancement quality compared to any single approach.
    
    \item \textbf{Classifier Guidance as a Control.} We adapt an Automatic Drum Transcription network into a noise-aware classifier and use it to guide the audio-only score, showing that the piano roll supplied only through classifier gradients does not substitute for conditioning the score network on MIDI.
    
    \item \textbf{MIDI-Free Inference.} We remove the ground-truth MIDI requirement at test time by transcribing the diffusion state online with the same time-conditional ADT network and handing over from the audio-only score to MiDiff at a switch time $t_{\text{sw}}$. Sweeping $t_{\text{sw}}$ reveals a clear optimum that recovers a large fraction of the symbolic-conditioning gain without any ground-truth piano roll.
    
    \item \textbf{Comprehensive Evaluation.} We evaluate our method using three complementary metrics: 1) Fréchet Audio Distance (FAD)~\cite{kilgour2018frechet}, 2) Mel Spectral Distance (MSD), and 3) Transient Timing Error (TTER). We demonstrate the results on the StemGMD dataset~\cite{stemgmd} with both validation and held-out test splits, and show that the gains are tied to the symbolic content: corrupted MIDI performs worse than the audio-only baseline. Ablation studies on velocity encoding and conditioning strategies provide further insights into the importance of authentic symbolic dynamics.
\end{enumerate}

\section{Background and Related Work}\label{sec:related_work}

\subsection{Symbolic Representation and Percussive Audio Processing}

MIDI~\cite{midi1983specification} represents music as discrete note-on and note-off events, specifying pitch, velocity, and timing. For percussion, General MIDI assigns a note number to each instrument (e.g., kick drum\,=\,36, snare drum\,=\,38), with strike intensity encoded by a velocity value ranging from 0 to 127. Learning-based systems commonly represent these events as a \emph{piano roll}, a two-dimensional grid of instrument classes and uniformly spaced time frames~\cite{huang2018music, oore2018time}. This representation is widely used in symbolic music generation and MIDI-conditioned audio synthesis~\cite{ibanez2023midi, he2025filling}. When temporally aligned with audio, the piano roll provides a noise-free description of the intended musical events, complementing the degraded audio signal.

Research on percussive audio encompasses source separation, synthesis, transcription, and mixing. HPSS~\cite{fitzgerald2010hpss} separates percussive components from polyphonic mixtures using median filtering. DrumGAN~\cite{nistal2020drumgan} and CRASH~\cite{rouard2021crash} synthesize one-shot drum samples using timbral descriptors and raw-waveform score-based diffusion, respectively, but generate audio from scratch rather than enhance existing recordings. Mezza~\textit{et al.}~\cite{stemgmd} introduced StemGMD for drum source separation, while Martinez Ramirez~\textit{et al.}~\cite{drumixing} investigated intelligent drum mixing. For automatic drum transcription, Vogl~\textit{et al.}~\cite{vogl2017drum} and Zehren~\textit{et al.}~\cite{zehren2023automatic} developed convolutional recurrent approaches to multi-class drum event detection (see Wu~\textit{et al.}~\cite{wu2018review} for a comprehensive review of automatic drum transcription). Most closely related to our work, a Cold Diffusion framework for percussive dereverberation~\cite{cold2026percussive} replaces stochastic noise with deterministic degradation, but does not incorporate symbolic conditioning.

\subsection{Score-Based Diffusion and Multimodal Conditioning}

Score-based diffusion~\cite{ho2020denoising,song2021scorebased} has become a leading approach to high-fidelity audio generation and enhancement. Lu~\textit{et al.}~\cite{lu2021diffuse} were among the first to apply diffusion models to speech enhancement. Richter~\textit{et al.}~\cite{sgmse} subsequently introduced SGMSE, operating in the complex STFT domain using an Ornstein--Uhlenbeck stochastic differential equation (SDE) and an NCSN++ U-Net backbone. This framework serves as the foundation for MiDiff. Related approaches, including CDiffuSE~\cite{lu2022conditional} and StoRM~\cite{lemercier2023storm}, further explored diffusion-based speech enhancement and dereverberation. 

For music, Moliner~\textit{et al.}~\cite{moliner2023solving} employed an unconditional diffusion prior to restore full-band recordings, while Hawthorne~\textit{et al.}~\cite{hawthorne2022multi} used spectrogram diffusion to synthesize audio from MIDI rather than restore existing recordings. The reverse diffusion process can be steered using classifier guidance~\cite{dhariwal2021diffusion} or classifier-free guidance~\cite{ho2021classifierfree}. In this work, we investigate the former. The benefits of incorporating complementary information into diffusion models have also been demonstrated in related tasks. Richter~\textit{et al.}~\cite{richter2023audio} showed that visual speech embeddings extracted using AV-HuBERT~\cite{shi2022avhubert} substantially improve the robustness of generative speech enhancement under severe degradation. LipVoicer~\cite{yemini2024lipvoicer} demonstrated that an auxiliary ASR classifier can guide the reverse diffusion process, while AudioLDM~\cite{liu2023audioldm} established the effectiveness of semantic conditioning in latent audio diffusion. These studies motivate the use of MIDI as an auxiliary information source and the guided sampling strategy proposed in MiDiff.

\subsection{Informed Enhancement and the Research Gap}

Score-informed source separation, pioneered by Ewert~\textit{et al.}~\cite{ewert2014score} and extended by Miron~\textit{et al.}~\cite{miron2017score}, uses aligned musical scores or MIDI transcriptions to constrain the separation of polyphonic audio. Meseguer-Brocal and Peeters~\cite{meseguer2020content} further incorporated time-aligned side information into a separation network using feature-wise linear modulation (FiLM). More generally, informed source separation~\cite{liutkus2012informed} exploits side information available at the encoding stage to guide source estimation. While these approaches aim to isolate individual sources from clean or mildly reverberant mixtures, MiDiff addresses informed \textit{enhancement} under severe non-additive distortions. Existing diffusion-based enhancement methods have primarily focused on speech, whereas MIDI-conditioned diffusion models have been used for audio \emph{synthesis} rather than \emph{enhancement}. To the best of our knowledge, no work has combined generative diffusion, MIDI conditioning, and drum-specific audio enhancement. MiDiff addresses this gap by incorporating symbolic information into diffusion-based restoration of degraded drum recordings.

\section{Problem Formulation}\label{sec:problem_formulation}

Let $\mathbf{y}\in\mathbb{C}^{F \times T}$ and $\mathbf{x}_0\in \mathbb{C}^{F \times T}$ denote the complex STFT spectrograms of degraded and clean drum audio signals, respectively, where $F$ is the number of frequency bins and $T$ is the number of time frames. We work in the complex STFT domain---rather than on magnitudes alone---because the percussive transients that dominate drum signals carry critical phase information that magnitude-only enhancement cannot reconstruct. Following SGMSE~\cite{sgmse}, we apply a magnitude-warping compression $\mathcal{T}(s) = \beta\,|s|^{\alpha}\,e^{i\angle s}$ with standard default parameters $\alpha = 0.5$ and $\beta = 0.15$ before feeding the spectrogram to the network; this tames the heavy-tailed magnitude distribution of drum spectra. The inverse transform $\mathcal{T}^{-1}$ is applied to the network output before the ISTFT.

The degraded observation $\mathbf{y}$ is generated from the clean signal $\mathbf{x}_0$ via a non-additive degradation process, defined formally as $\mathbf{y} = \mathcal{G}(\mathbf{x}_0)$. The operator $\mathcal{G}$ encompasses a sequence of non-linear distortions, which are detailed in \secref{subsec:degradation}.

The auxiliary conditioning signal $\mathbf{m}$ is represented as a temporally aligned MIDI piano-roll matrix $\mathbf{m}\in\mathbb{R}^{N_c\times T}$, where $N_c{=}9$ denotes the number of distinct drum instrument classes (Kick, Snare, Closed Hi-Hat, Open Hi-Hat, Ride, Crash, High Tom, Low-Mid Tom, High Floor Tom). Following the mapping convention by Mezza~\textit{et al.}~\cite{stemgmd}, these 9 classes consolidate the wide variety of raw MIDI note numbers (from both the General MIDI standard~\cite{generalmidi1991} and proprietary VST layouts~\cite{steinberg1999vst}) into a standardized vocabulary. The matrix entries $m_{k,t} \in [0, 1]$ represent 7-bit MIDI velocity values $v \in [0, 127]$ normalized to $[0, 1]$~\cite{he2025filling}, with zero indicating absence of note onset. Under these definitions, enhancement reduces to estimating $\mathbf{x}_0$ from $(\mathbf{y},\mathbf{m})$ by learning $p(\mathbf{x}_0 \mid \mathbf{y}, \mathbf{m})$.

\section{SGMSE Framework}\label{sec:proposed_method}

\subsection{Score-Based SDE Framework}
\label{subsec:sde_background}

Score-based generative models~\cite{song2021scorebased} unify DDPM~\cite{ho2020denoising} and score matching with Langevin dynamics (SMLD)~\cite{song2019generative} under a continuous-time SDE framework. The forward SDE gradually corrupts clean data $\mathbf{x}_0$ over continuous time $t \in [0, 1]$~\cite{gardiner1985handbook} via:
\begin{equation}\label{eq:forward_sde}
d\mathbf{x}_t = \mathbf{f}(\mathbf{x}_t, t)\, dt + g(t)\, d\mathbf{w},
\end{equation}
where $\mathbf{f}$ is the drift, $g(t)$ the diffusion coefficient, and $\mathbf{w}$ a standard Wiener process. Anderson's result~\cite{anderson1982reverse} shows that the time reversal is itself an SDE:
\begin{equation}\label{eq:reverse_sde}
d\mathbf{x}_t = \bigl[\mathbf{f}(\mathbf{x}_t, t) - g^2(t) \nabla_{\mathbf{x}_t}\log p_t(\mathbf{x}_t)\bigr]\, dt + g(t)\, d\bar{\mathbf{w}},
\end{equation}
where the intractable marginal score $\nabla_{\mathbf{x}_t}\log p_t(\mathbf{x}_t)$ is replaced by a network $s_\theta(\mathbf{x}_t, t)$ trained via denoising score matching (DSM)~\cite{vincent2011connection}:
\begin{multline}\label{eq:dsm}
\mathcal{L}_{\text{DSM}}(\theta) = \\\mathbb{E}_{t,\, \mathbf{x}_0,\, \mathbf{x}_t \mid \mathbf{x}_0}\!\left[\,\lambda(t)\,\bigl\| s_\theta(\mathbf{x}_t, t) - \nabla_{\mathbf{x}_t}\log p_{0t}(\mathbf{x}_t \mid \mathbf{x}_0) \bigr\|_2^2\,\right],
\end{multline}
where $\lambda(t) > 0$ is a time-dependent weighting function, and $p_{0t}(\mathbf{x}_t \mid \mathbf{x}_0)$ is the perturbation kernel (defined for our specific process in \secref{subsec:sgmse}).

\subsection{SGMSE: Enhancement via Ornstein-Uhlenbeck Process}
\label{subsec:sgmse}

For audio enhancement, SGMSE~\cite{sgmse} chooses an Ornstein-Uhlenbeck (OU) process drift~\cite{uhlenbeck1930theory, gardiner1985handbook} that transports the clean spectrum $\mathbf{x}_0$ toward the degraded observation $\mathbf{y}$:
\begin{equation}\label{eq:sgmse_forward}
d\mathbf{x}_t = \gamma (\mathbf{y} - \mathbf{x}_t) dt + g(t) d\mathbf{w}
\end{equation}
where $\gamma > 0$ is the OU stiffness parameter and $g(t)$ is the diffusion coefficient following the Variance Exploding (VE) exponential schedule $g(t) = \sigma_{\min}(\sigma_{\max}/\sigma_{\min})^t \sqrt{2\log(\sigma_{\max}/\sigma_{\min})}$. Because the drift is affine in $\mathbf{x}_t$, the perturbation kernel becomes Gaussian in closed form:
\begin{equation}\label{eq:sgmse_kernel}
p_{0t}(\mathbf{x}_t \mid \mathbf{x}_0, \mathbf{y}) = \mathcal{N}_{\mathbb{C}}\!\bigl(\mathbf{x}_t;\, \boldsymbol{\mu}(t),\, \sigma^2(t)\mathbf{I}\bigr),
\end{equation}
with mean and variance:
\begin{subequations}
\begin{align}
\boldsymbol{\mu}(t) &= e^{-\gamma t}\,\mathbf{x}_0 + \bigl(1 - e^{-\gamma t}\bigr)\,\mathbf{y}, \label{eq:sgmse_mean}\\
\sigma^2(t) &= \frac{\sigma_{\min}^2\left[\left(\frac{\sigma_{\max}}{\sigma_{\min}}\right)^{2t} - e^{-2\gamma t}\right] \log\left(\frac{\sigma_{\max}}{\sigma_{\min}}\right)}{\gamma + \log\left(\frac{\sigma_{\max}}{\sigma_{\min}}\right)}. \label{eq:sgmse_var}
\end{align}
\end{subequations}
The closed-form score target is given by
\begin{equation}
\nabla_{\mathbf{x}_t}\log p_{0t}(\mathbf{x}_t \mid \mathbf{x}_0, \mathbf{y}) = -\frac{\mathbf{x}_t - \boldsymbol{\mu}(t)}{\sigma^2(t)},
\end{equation}
so training requires no SDE simulation. To recover the enhanced spectrum $\mathbf{x}_0$, we solve the reverse-time SDE:

\begin{equation}\label{eq:reverse_ou}
d\mathbf{x}_t = \Bigl[\gamma(\mathbf{y}-\mathbf{x}_t) - g^2(t) s_\theta(\mathbf{x}_t, \mathbf{y}, t) \Bigr] dt 
+ g(t)\, d\bar{\mathbf{w}},
\end{equation}
parametrized by an NCSN++ U-Net backbone~\cite{song2021scorebased} and integrated numerically via a predictor-corrector sampler with $N$ discretization steps.

\subsection{Numerical Solution of the Reverse-Time SDE}
\label{subsec:numerical_reverse_sde}

Drawing a clean sample amounts to integrating \eqnref{eq:reverse_ou} numerically from $t=1$ back to $t=\epsilon$. We adopt the predictor--corrector (PC) sampler introduced by Song et~al.~\cite{song2021scorebased}.  The sampler alternates between a reverse-diffusion predictor step, which integrates \eqnref{eq:reverse_ou} across one interval of the discretized time grid, and an annealed Langevin dynamics corrector step $\mathbf{x} \gets \mathbf{x} + \delta\,s + \sqrt{2\delta}\,\mathbf{z}$, which refines the state at fixed $t$ with $\delta = 2\bigl(r\,\sigma(t)\bigr)^2$, corrector SNR $r$, and $\mathbf{z} \sim \mathcal{N}_{\mathbb{C}}(\mathbf{0},\mathbf{I})$. We write these two updates as $\textsc{Predictor}(\cdot)$ and $\textsc{Corrector}(\cdot)$ in what follows.

\section{MiDiff: MIDI-Conditioned Drum Audio Enhancement}\label{sec:midiff}

We extend the SGMSE framework to incorporate symbolic MIDI information by modeling the joint conditional score
\begin{equation}\label{eq:midi_score}
  s_\theta(\mathbf{x}_t, \mathbf{y}, \mathbf{m}, t) \approx \nabla_{\mathbf{x}_t} \log p_t(\mathbf{x}_t \mid \mathbf{y}, \mathbf{m}),
\end{equation}
where $\mathbf{m}$ is the synchronized MIDI piano-roll matrix defined in \secref{sec:problem_formulation}.
The central motivation is that MIDI provides an explicit, noise-free prior on \emph{which} drum instruments occur, \emph{when}, and \emph{how hard}, so that the score network need not infer instrument identity and strike dynamics from the heavily distorted acoustic observation $\mathbf{y}$ alone.

\subsection{MIDI Encoder}
\label{subsec:midi_encoder}

\begin{figure}[!t]
\centering
\includegraphics[width=0.8\linewidth]{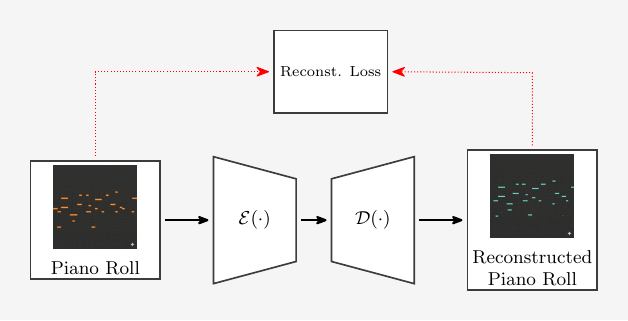}
\caption{MIDI encoder-decoder architecture: The raw piano roll $\mathbf{m}$ is encoded by $\mathcal{E}(\cdot)$ into a dense latent representation $\mathbf{z}$, and decoded by $\mathcal{D}(\cdot)$ to reconstruct $\hat{\mathbf{m}}$. Pretraining minimizes the squared reconstruction error.}

\label{fig:midi_autoencoder}
\end{figure}

The raw MIDI piano roll $\mathbf{m} \in [0,1]^{N_c \times T}$ is highly sparse and low-dimensional ($N_c$ drum classes), creating a severe semantic gap with the high-dimensional feature maps of the NCSN++ backbone. To overcome this mismatch, we employ a dedicated convolutional MIDI encoder $\mathcal{E}$ consisting of four 1D convolutional layers with Group Normalization and SiLU activation functions that progressively project the channel dimension from $N_c$ to latent dimension $C_M$ (in our implementation, $N_c{=}9$ drum classes and $C_M{=}128$).

As shown in~\figref{fig:midi_autoencoder}, the encoder is pretrained within an autoencoder framework minimizing reconstruction loss $\mathcal{L}_{\text{recon}} = \sum_{k,t} (m_{k,t} - \hat{m}_{k,t})^2$ where $\hat{\mathbf{m}} = \mathcal{D}(\mathcal{E}(\mathbf{m}))$. Once trained, the decoder $\mathcal{D}$ is discarded and the frozen encoder $\mathcal{E}$ serves as a deterministic feature extractor producing dense MIDI embeddings $\mathbf{M} = \mathcal{E}(\mathbf{m}) \in \mathbb{R}^{C_M \times T}$. Freezing the encoder stabilizes training, since the score network can rely on a fixed representation without contending with a co-evolving embedding space.

\subsection{Temporal Alignment}
\label{subsec:temporal_alignment}

A fundamental challenge in multimodal conditioning is resolving the temporal resolution mismatch across the network hierarchy. While the raw STFT spectrogram and the initial MIDI embedding $\mathbf{M} \in \mathbb{R}^{C_M \times T}$ share the same baseline frame count $T$, the NCSN++ U-Net backbone progressively downsamples intermediate acoustic representations. At depth level $\ell \in \{0, 1, \dots, L-1\}$, pooling and strided convolutions reduce the acoustic feature map dimensions to $\mathbf{h}^{(\ell)} \in \mathbb{R}^{C_\ell \times F_\ell \times T_\ell}$, where the temporal length scales as $T_\ell = T / 2^\ell$.
To supply temporally synchronized symbolic conditioning to every stage of the U-Net, we introduce a dedicated set of depth-specific Temporal Alignment modules, denoted as $\mathrm{TA}_\ell(\cdot)$. Rather than relying on simple linear interpolation or pooling, each $\mathrm{TA}_\ell$ is parameterized by a trainable 1D convolutional pipeline:
\begin{equation}\label{eq:temporal_alignment}
  \mathbf{M}^{(\ell)} = \mathrm{TA}_\ell(\mathbf{M}) = \mathrm{Conv1D}_{\text{refine}}\Bigl(\mathrm{SiLU}\bigl(\mathrm{Conv1D}_{\text{down}}(\mathbf{M})\bigr)\Bigr),
\end{equation}
where:
\begin{itemize}
    \item $\mathrm{Conv1D}_{\text{down}}$ uses an effective stride $S_\ell = 2^\ell$ and kernel size $k_\ell = 2^{\ell+1}$ with appropriate padding to downsample the sequence length from $T$ to $T_\ell$,
    \item $\mathrm{SiLU}(\cdot)$ is the sigmoid-weighted linear unit activation,
    \item $\mathrm{Conv1D}_{\text{refine}}$ is a unit-stride ($S=1$) convolution that refines the downsampled representation into the level-specific conditioning embedding $\mathbf{M}^{(\ell)} \in \mathbb{R}^{C_M \times T_\ell}$.
\end{itemize}
By learning dedicated downsampling kernels for each depth $\ell$, $\mathrm{TA}_\ell$ allows the network to summarize fine-grained musical transients (e.g., drum hit onsets) into broader, context-aware semantic features that directly align with the receptive field and resolution of the corresponding U-Net layer.

\subsection{MIDI Conditioning Methods}
\label{subsec:midi_conditioning}

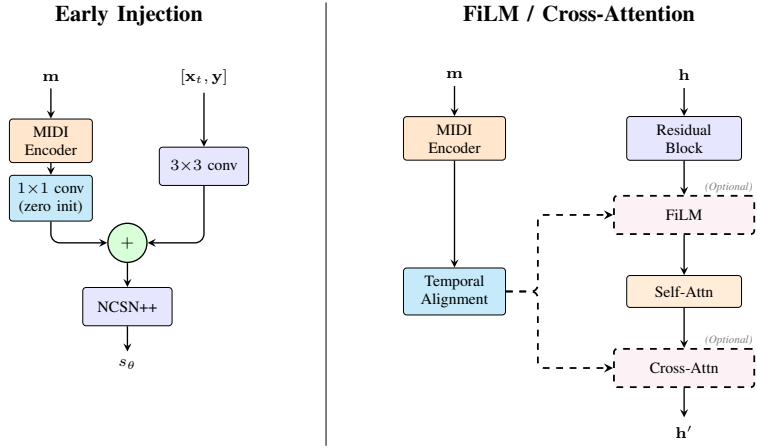
\begin{figure*}[!t]
\centering
\resizebox{0.55\textwidth}{!}{%
\begin{tabular}{@{}c@{\hspace{12mm}}|@{\hspace{12mm}}c@{}}
\textbf{Early Injection} & \textbf{FiLM / Cross-Attention} \\[6pt]
\begin{tikzpicture}[scale=1, transform shape, font=\scriptsize, baseline=(current bounding box.north),
  unetbox/.style = {draw, rounded corners=1.5pt, minimum width=14mm, minimum height=6mm, align=center, fill=blue!10},
  plusbox/.style = {circle, draw, minimum size=5mm, align=center, fill=green!15, semithick},
  encoderbox/.style = {draw, rounded corners=1.5pt, minimum width=13mm, minimum height=5mm, align=center, fill=orange!20},
  arrow/.style = {->, >=stealth, semithick}
]
\node (m_raw) at (-12mm,20mm) {$\mathbf{m}$};
\node (xt) at (12mm,20mm) {$[\mathbf{x}_t, \mathbf{y}]$};
\node[encoderbox] (encoder) at (-12mm,10mm) {MIDI\\Encoder};
\node[encoderbox, fill=cyan!20] (zconv) at (-12mm,1mm) {$1{\times}1$ conv\\(zero init)};
\node[unetbox] (conv0) at (12mm,6mm) {$3{\times}3$ conv};
\node[plusbox] (plus2) at (0,-6mm) {$+$};
\node[unetbox] (unet) at (0,-16mm) {NCSN++};
\node (out) at (0,-25mm) {$s_\theta$};

\draw[arrow] (m_raw) -- (encoder);
\draw[arrow] (encoder) -- (zconv);
\draw[arrow, rounded corners=4pt] (zconv.south) -- (zconv.south |- plus2.west) -- (plus2.west);
\draw[arrow] (xt) -- (conv0);
\draw[arrow, rounded corners=4pt] (conv0.south) -- (conv0.south |- plus2.east) -- (plus2.east);
\draw[arrow] (plus2) -- (unet);
\draw[arrow] (unet) -- (out);
\end{tikzpicture}
&
\begin{tikzpicture}[scale=1, transform shape, font=\scriptsize, baseline=(current bounding box.north),
  resbox/.style = {draw, rounded corners=1.5pt, minimum width=18mm, minimum height=5mm, align=center, fill=blue!10},
  selfbox/.style = {draw, rounded corners=1.5pt, minimum width=18mm, minimum height=5mm, align=center, fill=orange!15},
  condbox/.style = {draw, dashed, thick, rounded corners=1.5pt, minimum width=22mm, minimum height=6mm, align=center, fill=purple!5},
  encoderbox/.style = {draw, rounded corners=1.5pt, minimum width=16mm, minimum height=5mm, align=center, fill=orange!20},
  alignbox/.style = {draw, rounded corners=1.5pt, minimum width=16mm, minimum height=5mm, align=center, fill=cyan!20},
  arrow/.style = {->, >=stealth, semithick},
  optarrow/.style = {->, >=stealth, dashed, thick}
]
\node (h_in) at (0, 28mm) {$\mathbf{h}$};
\node[resbox] (res1) at (0, 18mm) {Residual \\ Block};

\node[condbox] (film) at (0, 6mm) {FiLM};
\node[anchor=south east, font=\tiny\itshape, text=gray, inner sep=1pt] at (film.north east) {(Optional)};

\node[selfbox] (self) at (0, -6mm) {Self-Attn};

\node[condbox] (cross) at (0, -18mm) {Cross-Attn};
\node[anchor=south east, font=\tiny\itshape, text=gray, inner sep=1pt] at (cross.north east) {(Optional)};

\node (h_out) at (0, -28mm) {$\mathbf{h'}$};

\node (m_raw) at (-36mm, 28mm) {$\mathbf{m}$};
\node[encoderbox] (midi_encoder) at (-36mm, 18mm) {MIDI\\Encoder};
\node[alignbox] (temporal_align) at (-36mm, -6mm) {Temporal\\Alignment};

\draw[arrow] (h_in) -- (res1);
\draw[arrow] (res1) -- (film);
\draw[arrow] (film) -- (self);
\draw[arrow] (self) -- (cross);
\draw[arrow] (cross) -- (h_out);

\draw[arrow] (m_raw) -- (midi_encoder);
\draw[arrow] (midi_encoder) -- (temporal_align);

\draw[optarrow, rounded corners=3pt] (temporal_align.east) -- ++(5mm,0) |- (film.west);
\draw[optarrow, rounded corners=3pt] (temporal_align.east) -- ++(5mm,0) |- (cross.west);

\end{tikzpicture}
\end{tabular}
}
\caption{\textbf{Left:} Early Injection mechanism---encodes raw piano roll $\mathbf{m}$, projects the embedding with a zero-initialized $1\times1$ convolution and adds it to the output of the first NCSN++ convolution applied to $[\mathbf{x}_t,\mathbf{y}]$ (Eq.~\eqnref{eq:early_injection}). \textbf{Right:} Hierarchical conditioning blocks---acoustic features $\mathbf{h}$ pass through residual blocks, optional FiLM modulation, self-attention, and optional cross-attention blocks driven by scale-aligned MIDI embeddings.}
\label{fig:midi_conditioning}
\end{figure*}

We explore three distinct conditioning strategies to integrate MIDI priors into the NCSN++ score network (\figref{fig:midi_conditioning}):

\noindent\textbf{Early Injection.} The four-channel input $[\mathbf{x}_t, \mathbf{y}]$ (real and imaginary parts of both spectrograms) first passes through the initial $3\times3$ convolution of NCSN++, which lifts it to $C_0$ feature channels. The full-resolution MIDI embedding $\mathbf{M}^{(0)}$ is mapped to the same width by a $1\times1$ convolution $\mathbf{W}_{\text{inj}}\in\mathbb{R}^{C_0\times C_M}$, broadcast across the frequency axis and added to these first-layer features:
\begin{equation}\label{eq:early_injection}
  \mathbf{h}^{(0)} = \mathrm{Conv}_{3\times3}\bigl([\mathbf{x}_t,\mathbf{y}]\bigr) + \mathbf{W}_{\text{inj}}\,\mathbf{M}^{(0)} .
\end{equation}
Since $\mathbf{W}_{\text{inj}}$ and its bias are zero-initialized, the injection starts as the identity, as in ControlNet-style zero convolutions; adding a projection after the first linear layer is equivalent to concatenating $\mathbf{M}^{(0)}$ to the network input.

\noindent\textbf{FiLM (Feature-wise Linear Modulation).} Following Perez~\textit{et al.}~\cite{perez2018film}, intermediate spectral-temporal feature maps $\mathbf{h} \in \mathbb{R}^{C \times F \times T_\ell}$ at each U-Net layer are modulated by affine parameters predicted from $\mathbf{M}^{(\ell)}$, where $F$ is the number of STFT frequency bins and $T_\ell$ is the temporal length of features at encoder depth $\ell \in \{1,2,3,4\}$:

\begin{subequations}

\begin{align}
    [\mathbf{a}, \mathbf{b}] &= \text{Conv1D}_{1\times 1}(\mathbf{M}^{(\ell)})  \\
    \text{FiLM}(\mathbf{h}) &= (1 + \mathbf{a}) \odot \mathbf{h} + \mathbf{b} \label{eq:film}
\end{align}
\end{subequations}
where $\mathbf{a}, \mathbf{b} \in \mathbb{R}^{C \times 1 \times T_\ell}$ modulate channel scale and shift across frequency bins $F$. The $(1+\mathbf{a})$ term with zero-initialized projection weights ensures FiLM blocks default to an identity transform at initialization, stabilizing early training. FiLM is inserted after each NCSN++ residual block to enable hierarchical modulation across different semantic scales.

\noindent\textbf{Cross-Attention.} Following Vaswani~\textit{et al.}~\cite{vaswani2017attention}, we insert cross-attention blocks in which the U-Net acoustic feature maps $\mathbf{h} \in \mathbb{R}^{C \times F \times T_\ell}$ act as queries and the level-aligned MIDI embeddings $\mathbf{M}^{(\ell)} \in \mathbb{R}^{C_M \times T_\ell}$ serve as keys and values, with $\ell$ denoting the U-Net encoder depth and $T_\ell$ the corresponding temporal resolution (both as defined in \secref{subsec:temporal_alignment}). After reshaping to sequence form, the queries, keys, and values are obtained via learned linear projections:
\begin{equation}\label{eq:cross_attn_proj}
  \mathbf{Q} = \mathbf{h}\mathbf{W}^Q, \quad
  \mathbf{K} = \mathbf{M}^{(\ell)}\mathbf{W}^K, \quad
  \mathbf{V} = \mathbf{M}^{(\ell)}\mathbf{W}^V,
\end{equation}
where $\mathbf{W}^Q \in \mathbb{R}^{C \times d_k}$, $\mathbf{W}^K, \mathbf{W}^V \in \mathbb{R}^{C_M \times d_k}$ and $d_k = C / N_h$ for $N_h$ attention heads. Standard scaled dot-product attention~\cite{vaswani2017attention} is then applied, allowing each acoustic location to dynamically attend over all MIDI time steps at the downsampled U-Net depths $\ell \in \{1,2,3,4\}$ (omitting the highest-resolution layer $\ell=0$).

Our unified architecture, \textbf{MiDiff}, integrates all three conditioning strategies simultaneously within the NCSN++ backbone: the MIDI embedding is added at the input (early injection), each residual block is FiLM-modulated, and a cross-attention block is inserted after each self-attention layer.
The individual contribution of each pathway is quantified via an ablation study (in~\secref{sec:ablation}), where each strategy is evaluated alone and in combination to assess its effect on enhancement quality.

\subsection{Training Objective}
\label{subsec:training_objective}

Given the closed-form perturbation kernel in \eqnref{eq:sgmse_kernel} with parameters defined in \eqnref{eq:sgmse_mean}--\eqnref{eq:sgmse_var}, we train the MIDI-conditioned score network by solving~\cite{sgmse}:

\begin{equation}\label{eq:midiff_loss}
\begin{aligned}
\hat{\theta} = \argmin_\theta \mathbb{E}_{t, (\mathbf{x}_0, \mathbf{y}, \mathbf{m}), \mathbf{z}, \mathbf{x}_t \mid (\mathbf{x}_0, \mathbf{y})} \Biggl[ \biggl\| &s_\theta(\mathbf{x}_t, \mathbf{y}, \mathbf{m}, t) \\
&+ \frac{\mathbf{z}}{\sigma(t)} \biggr\|_2^2 \Biggr].
\end{aligned}
\end{equation}
This objective is directly derived from the Denoising Score Matching (DSM) loss in \eqnref{eq:dsm} by substituting the analytical score $\nabla_{\mathbf{x}_t}\log p_{0t}(\mathbf{x}_t \mid \mathbf{x}_0, \mathbf{y}) = -{\mathbf{z}}/{\sigma(t)}$ and setting the weighting function $\lambda(t) = \sigma^2(t)$. Here, the expectation is over $t$ sampled from a uniform distribution $\mathcal{U}[\epsilon, 1]$, data triplets $(\mathbf{x}_0, \mathbf{y}, \mathbf{m})$ sampled from the empirical training distribution $\mathcal{P}$, and standard complex Gaussian noise $\mathbf{z} \sim \mathcal{N}_{\mathbb{C}}(\mathbf{0}, \mathbf{I})$ which parameterizes the perturbed audio as $\mathbf{x}_t = \boldsymbol{\mu}(t) + \sigma(t)\mathbf{z}$.

MiDiff is trained with the \emph{always present} piano roll, without conditioning dropout, since it is always conditioned on MIDI at inference.

\subsection{Conditioned Reverse Sampling}
\label{subsec:midi_sampling}

At test time, MiDiff reuses the predictor--corrector sampler (detailed in \secref{subsec:numerical_reverse_sde}) without altering its predictor or corrector updates. The trajectory is initialized at the terminal distribution of the forward SDE at time $t=1$, which is approximately a Gaussian centered on the noisy observation, i.e.~$\mathbf{x}_1 \sim \mathcal{N}_{\mathbb{C}}(\mathbf{y}, \sigma^2(1)\mathbf{I})$, and is then integrated backward down to $t=\epsilon$. The only change from the base sampler is that every score evaluation is replaced by the MIDI-conditioned score $s_\theta(\mathbf{x}_t,\mathbf{y},\mathbf{m},t)$, steering every reverse step. Algorithm~\ref{alg:hybrid_pc} in \secref{subsec:hybrid_sampling} states the resulting procedure in full; setting $t_{\text{sw}}{=}1$ there and substituting the ground-truth piano roll $\mathbf{m}$ for the online transcription $\hat{\mathbf{m}}$ recovers exactly the sampler described here.

\subsection{Classifier Guidance via Audio-to-MIDI}
\label{subsec:classifier_guidance}
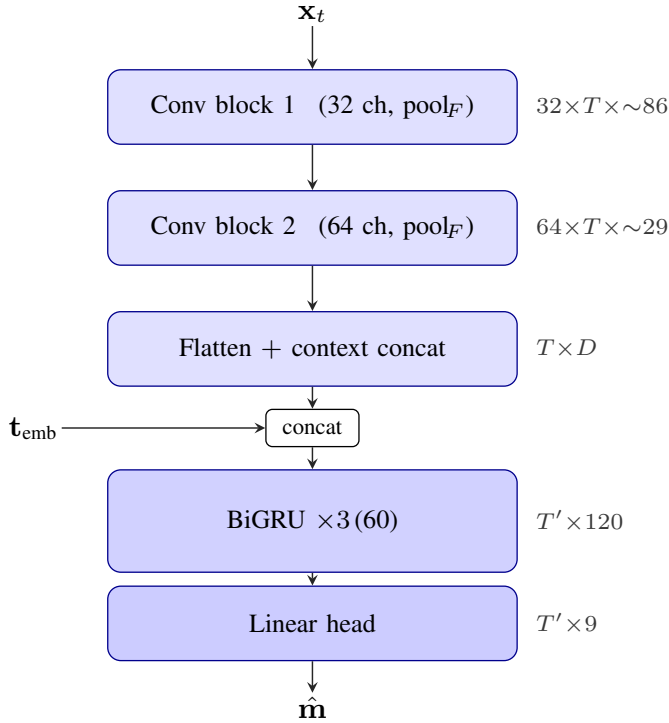
\begin{figure}[ht]
\centering
\resizebox{\linewidth}{!}{%
\begin{tikzpicture}[
  font=\footnotesize,
  >=stealth,
  every node/.style={inner sep=2pt},
  box/.style={draw=blue!55!black, line width=0.4pt, rounded corners=4pt,
              fill=blue!12, minimum height=8mm, minimum width=44mm, align=center},
  rnn/.style={draw=blue!55!black, line width=0.4pt, rounded corners=4pt,
              fill=blue!18, minimum height=11mm, minimum width=44mm, align=center},
  head/.style={draw=blue!55!black, line width=0.4pt, rounded corners=4pt,
               fill=blue!20, minimum height=8mm, minimum width=44mm, align=center},
  sum/.style={draw=black, line width=0.4pt, rounded corners=2pt, fill=white,
              inner sep=1pt, minimum height=4mm, minimum width=10mm, font=\scriptsize},
  arr/.style={->, line width=0.45pt, draw=black!85},
  hlabel/.style={font=\small, inner sep=1pt},
  shp/.style={font=\scriptsize\itshape, gray!45!black}
]
  \node[hlabel] (xin) at (0, 0)                 {$\mathbf{x}_t$};
  \node[box]   (c1)   at (0,-1.0)               {Conv block 1 \, (32 ch, pool$_{\!F}$)};
  \node[box]   (c2)   at (0,-2.3)               {Conv block 2 \, (64 ch, pool$_{\!F}$)};
  \node[box]   (flat) at (0,-3.6)               {Flatten $+$ context concat};
  \node[sum]   (cat)  at (0,-4.45)              {concat};
  \node[rnn]   (gru)  at (0,-5.45)              {BiGRU $\times 3$\,(60)};
  \node[head]  (lin)  at (0,-6.55)              {Linear head};
  \node[hlabel](yout) at (0,-7.45)              {$\hat{\mathbf{m}}$};

  \node[hlabel] (temb) at (-3.0,-4.45) {$\mathbf{t}_{\text{emb}}$};

  \draw[arr] (xin)  -- (c1);
  \draw[arr] (c1)   -- (c2);
  \draw[arr] (c2)   -- (flat);
  \draw[arr] (flat) -- (cat);
  \draw[arr] (cat)  -- (gru);
  \draw[arr] (gru)  -- (lin);
  \draw[arr] (lin)  -- (yout);

  \draw[arr] (temb.east) -- (cat.west);

  \node[shp, anchor=west] at ($(c1.east)+(0.12,0)$)   {$32{\times}T{\times}{\sim}86$};
  \node[shp, anchor=west] at ($(c2.east)+(0.12,0)$)   {$64{\times}T{\times}{\sim}29$};
  \node[shp, anchor=west] at ($(flat.east)+(0.12,0)$) {$T{\times}D$};
  \node[shp, anchor=west] at ($(gru.east)+(0.12,0)$)  {$T'{\times}120$};
  \node[shp, anchor=west] at ($(lin.east)+(0.12,0)$)  {$T'{\times}9$};
\end{tikzpicture}%
}
\caption{Time-conditional ADT classifier architecture: convolutional encoder, context concatenation, BiGRU stack and linear head producing per-frame onset-probability logits $\hat{\mathbf{m}}$---the MIDI \emph{inferred} from the noisy audio $\mathbf{x}_t$, distinct from the ground-truth piano roll $\mathbf{m}$. The 128-dimensional diffusion-time embedding vector $\mathbf{t}_{\text{emb}}$ is concatenated into the recurrent input at the \textit{concat} node.}

\label{fig:adt_architecture}
\end{figure}

Inspired by LipVoicer~\cite{yemini2024lipvoicer}, which guides speech diffusion using an auxiliary ASR classifier, we adapt the ADTOF network~\cite{zehren2023automatic} into a time-conditioned classifier that models $p_\phi(\mathbf{m} \mid \mathbf{x}_t, t)$. Given the noisy diffusion state $\mathbf{x}_t$, the classifier estimates onset probabilities $\hat{\mathbf{m}}$, while the ground-truth MIDI $\mathbf{m}$ serves as the target condition. The gradient of the log-probability of the ground-truth MIDI is then used to guide the diffusion process. Since standard ADT models are trained on clean audio and lack a mechanism to condition on diffusion time $t$, we extend the ADTOF backbone with a parallel $t$-conditioning pathway, as detailed below. 

The backbone, sketched in \figref{fig:adt_architecture}, takes the noisy diffusion state $\mathbf{x}_t$, represented as a two-channel real/imaginary tensor. Two convolutional blocks of $\{32,64\}$ channels apply $3\!\times\!3$ convolutions, each followed by a max-pool of width 3 along the frequency axis only; this compresses the spectral dimension while preserving the per-frame temporal resolution required for onset prediction. The tensor is flattened, and a context layer concatenates a short window of adjacent frames before the recurrent stage. Three bidirectional GRUs of 60 hidden units model temporal structure, and a linear head emits the per-frame logits $\hat{\mathbf{m}}$. Diffusion time $t$ enters through a parallel MLP pathway ($1{\to}128{\to}128$ with SiLU), concatenated at the GRU input, allowing the network to adapt its predictions to the current noise level.

Critically, the classifier is trained on the same OU--VE perturbed marginals seen during score network training: $t \sim \mathcal{U}(\epsilon,1)$, $\mathbf{x}_t \sim p_{0t}(\cdot\mid\mathbf{x}_0,\mathbf{y})$, with per-class binary cross-entropy loss between the prediction $\hat{\mathbf{m}}$ and the true MIDI $\mathbf{m}$. Training the classifier explicitly on noisy samples $\mathbf{x}_t$ conditioned on the diffusion time $t$ ensures that its gradient remains informative and accurate across the entire reverse diffusion trajectory, even at high noise levels. The guided score combines the unconditional score with a weighted classifier gradient:
\begin{equation}\label{eq:cg_score}
\hat{\mathbf{s}}_\psi(\mathbf{x}_t, \mathbf{y}, \mathbf{m}, t) = \mathbf{s}_\psi(\mathbf{x}_t, \mathbf{y}, t) + w_{\text{CG}} \cdot \nabla_{\mathbf{x}_t} \log p_\phi(\mathbf{m} \mid \mathbf{x}_t, t),
\end{equation}
where $w_{\text{CG}} \ge 0$ controls classifier guidance strength. The guidance term is added to the \emph{unconditional} score $\mathbf{s}_\psi(\mathbf{x}_t,\mathbf{y},t)$ of the audio-only SGMSE model (\secref{subsec:baselines}), which has no MIDI input. The piano roll $\mathbf{m}$ therefore reaches the sampler only through the classifier gradient, and $w_{\text{CG}}{=}0$ reduces Eq.~\eqnref{eq:cg_score} to plain SGMSE. This isolates whether symbolic information introduced only at inference time, externally to the score network, can replace explicit MIDI conditioning of the network. Here, $\phi$ denotes the classifier parameters in \figref{fig:adt_architecture}. Note that $\log p_\phi(\mathbf{m} \mid \mathbf{x}_t, t)$ corresponds to the negative cross-entropy between the classifier's prediction $\hat{\mathbf{m}}$ and the target $\mathbf{m}$. By taking the gradient of this term with respect to $\mathbf{x}_t$, we compute the direction in the noisy audio space that most increases the likelihood of the given MIDI, effectively steering the generative trajectory toward the desired drum pattern.

\subsection{Inference without Ground-Truth MIDI: Hybrid Sampling}
\label{subsec:hybrid_sampling}

The preceding experiments assume that a temporally aligned piano roll $\mathbf{m}$ is available at test time. This assumption holds when the performance is recorded using electronic pads or sequenced in a DAW, but not when only an audio recording is available. In the latter case, MiDiff must rely on MIDI transcribed from the recording itself. We obtain this transcription \emph{online} using the noise-level-aware ADT network described in \secref{subsec:classifier_guidance}, now employed as a transcriber rather than to provide guidance gradients.

At each reverse step, the network transcribes the current diffusion state:
\begin{equation}\label{eq:online_adt}
\hat{\mathbf{m}}_t = \mathrm{ADT}_\phi(\mathbf{x}_t, t),
\end{equation}
and the resulting drum roll is used as the condition for that step. The onset probabilities are converted to a piano roll in the format MiDiff was trained on: per-class peak picking with class-specific thresholds (minimum spacing of three frames), each detected onset held for 13 frames at a constant normalized velocity of $0.43$, since the transcriber predicts onsets but not velocities. Because the same $t$-conditioned network is already trained on the OU--VE perturbed marginals, no additional model is required.

Two failure modes make na\"{i}ve online conditioning unattractive. First, the transcriber is least reliable at high noise levels, where $\mathbf{x}_t$ is still close to the degraded observation $\mathbf{y}$ and the onset activations collapse (\figref{fig:adt_progression}). Second, MiDiff is trained exclusively on ground-truth piano rolls and therefore never saw erroneous conditioning during training, so a corrupted $\hat{\mathbf{m}}_t$ injected early in the trajectory is propagated through the remaining reverse steps. We therefore split the reverse trajectory at a switch time $t_{\text{sw}} \in [0,1]$ and let the audio-only SGMSE score $s_\psi(\mathbf{x}_t,\mathbf{y},t)$, namely the baseline model of \secref{subsec:baselines}, carry the high-noise portion and hand over to MiDiff for the remainder, with the value of $t_{\text{sw}}$ examined in \secref{subsec:hybrid_results}:
\begin{equation}\label{eq:hybrid_score}
s^{\text{hyb}}(\mathbf{x}_t,\mathbf{y},t) =
\begin{cases}
s_\psi(\mathbf{x}_t,\mathbf{y},t), & t > t_{\text{sw}}, \\
s_\theta(\mathbf{x}_t,\mathbf{y},\hat{\mathbf{m}}_t,t), & t \le t_{\text{sw}}.
\end{cases}
\end{equation}
Both models share the same OU--VE SDE and complex STFT representation, so the state is handed over unchanged and the same predictor--corrector sampler is used on both sides of the switch; only the score evaluation differs. Algorithm~\ref{alg:hybrid_pc} summarizes the procedure: the transcription $\hat{\mathbf{m}}$ is computed once per step and shared by the corrector and the predictor, whereas the score is re-evaluated at the corrected state before the predictor update. The two extremes recover the individual models: $t_{\text{sw}}{=}0$ is SGMSE alone, since the reverse process terminates at $\epsilon > 0$ and the switch is never reached, while $t_{\text{sw}}{=}1$ is MiDiff conditioned on online ADT transcriptions at all steps. Note that $t_{\text{sw}}$ is a pure inference-time knob: neither network is retrained, and the ground-truth-MIDI results of \secref{subsec:midi_sampling} are unaffected.

\begin{algorithm}[ht]
\caption{Hybrid SGMSE\,$\rightarrow$\,MiDiff PC sampling with online ADT conditioning.}
\label{alg:hybrid_pc}
\begin{algorithmic}[1]
\REQUIRE $\mathbf{y}$; scores $s_\psi$ (audio-only) and $s_\theta$ (MIDI-conditioned); transcriber $\mathrm{ADT}_\phi$; switch time $t_{\text{sw}}$
\ENSURE enhanced estimate $\hat{\mathbf{x}}$
\STATE $\mathbf{x} \sim \mathcal{N}_{\mathbb{C}}(\mathbf{y},\,\sigma^2(1)\mathbf{I})$ \COMMENT{OU--VE prior centered on $\mathbf{y}$}
\FOR{$t = t_1,\dots,t_N$ \textbf{uniformly spaced in} $[1,\epsilon]$}
  \IF{$t > t_{\text{sw}}$}
    \STATE $s(\cdot) \gets s_\psi(\cdot,\mathbf{y},t)$ \COMMENT{SGMSE: no MIDI}
  \ELSE
    \STATE $\hat{\mathbf{m}} \gets \mathrm{ADT}_\phi(\mathbf{x},t)$ \COMMENT{transcribe current state}
    \STATE $s(\cdot) \gets s_\theta(\cdot,\mathbf{y},\hat{\mathbf{m}},t)$ \COMMENT{MiDiff}
  \ENDIF
  \STATE $\mathbf{x} \gets \textsc{Corrector}(\mathbf{x},s(\mathbf{x}),t)$ \COMMENT{annealed Langevin}
  \STATE $\mathbf{x} \gets \textsc{Predictor}(\mathbf{x},\mathbf{y},s(\mathbf{x}),t)$ \COMMENT{score re-evaluated at corrected $\mathbf{x}$}
\ENDFOR
\RETURN $\hat{\mathbf{x}} \gets \text{ISTFT}\bigl(\mathcal{T}^{-1}(\mathbf{x})\bigr)$
\end{algorithmic}
\end{algorithm}

\section{Experimental Setup}\label{sec:experimental_setup}

\subsection{StemGMD Dataset}
\label{subsec:datasets}

\begin{figure}[!t]
  \centering
  \includegraphics[width=0.85\linewidth]{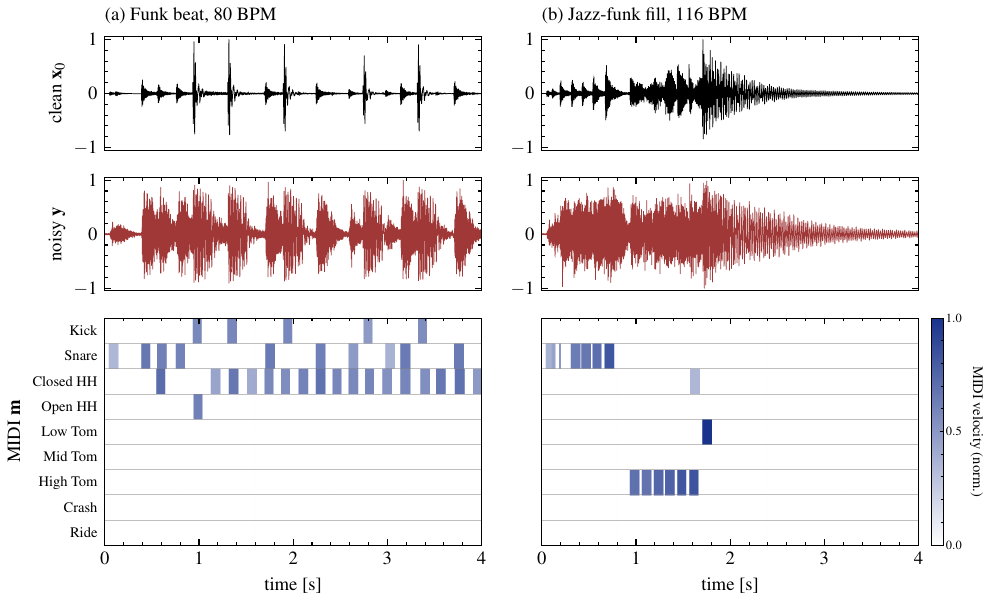}
  \caption{Two StemGMD examples (4\,s each). From top: clean reference $\mathbf{x}_0$, degraded input $\mathbf{y}$, and piano-roll $\mathbf{m}$ (one row per drum class, cell intensity $\propto$ MIDI velocity). Panels share their time axis---every audio transient matches a non-zero cell in $\mathbf{m}$.}
  \label{fig:midi_audio_sync}
\end{figure}

Training a paired enhancement model requires triplets $(\mathbf{x}_0, \mathbf{y}, \mathbf{m})$ with sample-accurate alignment. Building such a corpus from real recordings is impractical: simultaneously capturing studio-grade and low-quality signals from the same performance introduces unavoidable latency, level, and positional differences. We use the StemGMD dataset~\cite{stemgmd}, a 1224-hour corpus of synthesized drum stems comprising expressive, microtimed human drum performances captured by ten drummers across ten acoustic drum kit libraries, with original velocity and microtiming preserved in MIDI.\footnote{This research utilizes the open StemGMD dataset in compliance with dataset licensing terms.}

We selected StemGMD over purely algorithmic multi-instrument corpora such as Slakh~\cite{manilow2019cutting} because it contains authentic human performances from the Groove MIDI Dataset~\cite{gillick2019groove} (captured as MIDI via electronic drum kits) that preserve natural microtiming and mechanical-strike dynamics. To generate the audio, these human-played MIDI performances are mapped to a canonical 9-piece kit (Kick, Snare, Closed Hi-Hat, Open Hi-Hat, High Tom, Low-Mid Tom, High Floor Tom, Crash, Ride) and independently rendered as 16-bit/44.1\,kHz WAV files. This rendering is performed through ten different high-fidelity Logic Pro X acoustic drum kit libraries, providing varied timbres, room acoustics, and overhead-mic characteristics.
The clean reference mixture $\mathbf{x}_0 = \sum_{k=1}^{9} s_k$ is obtained by summing the nine isolated stems. Because these stems are synthesized directly from the MIDI rather than being acoustically recorded together in a physical room, $\mathbf{x}_0$ is entirely free of microphone bleed and environmental artifacts by construction. The same construction makes the conditioning an oracle: $\mathbf{m}$ is the generative cause of $\mathbf{x}_0$, so its alignment with the target is exact, a condition that real recordings only approximate (\secref{subsec:limitations}). Finally, the noisy observation $\mathbf{y}$ is generated offline from $\mathbf{x}_0$ by the degradation pipeline described below, once per clip, so every model sees the same fixed degraded input. This synthetic pipeline yields properties that no real corpus can offer simultaneously: an exact, clean reference for every example (enabling FAD evaluation) and a known, controllable degradation operator.

\figref{fig:midi_audio_sync} illustrates the alignment between clean audio, degraded audio, and MIDI piano-roll conditioning for two StemGMD examples.

\begin{table}[!t]
  \caption{StemGMD Dataset Splits (16\,kHz Mono).}
  \label{tab:dataset_splits}
  \centering
  \small
  \begin{tabularx}{\linewidth}{@{}l r X@{}}
    \toprule
    \textbf{Split} & \textbf{Duration} & \textbf{Source / Configuration} \\
    \midrule
    Train & 16.08\,h & \texttt{drummer1/session1} (low-data setup) \\
    Validation & 6.9\,min & 10 \texttt{eval\_session} recordings, 10 styles \\
    Test & 1.16\,h & \texttt{drummer1/eval\_session} (100 clips) \\
    \bottomrule
  \end{tabularx}
\end{table}

\tabref{tab:dataset_splits} summarizes the dataset splits. Training uses only the \texttt{drummer1/session1} subset (16.08\,h), a deliberately data-constrained setting in which paired data is scarce. The test split is the \texttt{drummer1/eval\_session} folder (100 clips, 1.16\,h): recordings disjoint from training, but by the same drummer and rendered with the same ten kit libraries. The validation split consists of 10 \texttt{eval\_session} recordings, one per drum style (6.9\,min), with a separately drawn degradation, and is used for epoch-wise FAD monitoring and checkpoint selection.

\subsection{Realistic Degradation Pipeline}
\label{subsec:degradation}

To simulate a low-quality home recording, clean drum mixtures are processed offline through a three-stage non-additive pipeline, with every free parameter drawn independently for each clip. The ordering follows the physical signal path (room, capsule, electronics), and no single factor dominates, so the model cannot succeed by inverting one filter:

\begin{enumerate}
    \item \textbf{Room Reverberation:} 
The signal is convolved with a synthetic impulse response generated from Gaussian noise (mean 1, standard deviation 0.5) modulated by an exponential envelope $e^{-t/(0.4\,T_d)}$, where $T_d \sim \mathcal{U}[0.12, 0.35]$\,s denotes the response duration. The impulse response is normalized to unit power, and the reverberant signal is mixed with the dry signal at a wet ratio drawn from $\mathcal{U}[0.08, 0.22]$. A single discrete echo, with a delay drawn from $\mathcal{U}[25, 70]$\,ms and a gain drawn from $\mathcal{U}[0.12, 0.35]$, is then added to emulate a strong early reflection. Finally, the resulting signal is peak-normalized.

    \item \textbf{Electret Microphone Coloration:} A three-filter model comprising a 4th-order Butterworth high-pass (cut-on $\sim\mathcal{U}[100, 250]$\,Hz); a second-order resonant band-pass centred at $f_p \sim \mathcal{U}[3, 5]$\,kHz with $Q{=}2$ and gain $1+g/10$, $g\sim\mathcal{U}[3, 8]$, which concentrates the signal in the upper midrange (the ``honk'' of small capsules) and strongly attenuates the low end; and a 4th-order Butterworth low-pass (cut-off $\sim\mathcal{U}[3.2, 7.5]$\,kHz).
    \item \textbf{Nonlinear Saturation \& Noise:} Preamp saturation is modeled by a soft $\tanh$ clipper $x\mapsto\tanh(\alpha\,x)$ with drive $\alpha \sim \mathcal{U}[1.0, 5.0]$, followed by peak normalization. Additive pink ($1/f$) noise is then scaled to hit a target signal-to-noise ratio $\text{SNR} \sim \mathcal{U}[45, 65]$\,dB.
\end{enumerate}

We deliberately focus on single-instrument drum audio restoration under acoustic and recording-chain degradations (reverberation, microphone coloration, clipping, and noise), rather than multi-instrument bleed or source-separation artifacts. This allows us to isolate the contribution of symbolic MIDI priors to percussive enhancement without confounding errors from source separation. The combined degradations are severe: after RMS normalization, the degraded signal is a median of $49$\,dB below the clean signal in the 20--60\,Hz band and $34$\,dB below it in the 60--120\,Hz band, nearly eliminating the kick fundamental. Conversely, in the 1--8\,kHz range, the degraded signal exceeds the clean signal by $14$--$16$\,dB, resulting in an input dominated by cymbals and saturation products.

\subsection{Implementation Details}
\label{subsec:implementation_details}

All audio signals are resampled to 16\,kHz. STFT spectrograms are computed using a 510-sample Hann window ($\approx$31.8\,ms) and a hop size of 128 samples ($\approx$8.0\,ms), yielding $F{=}256$ complex frequency bins matching the NCSN++ downsampling geometry while preserving temporal resolution for transient detection.

Following the SGMSE framework~\cite{sgmse}, the forward OU-VE SDE is parameterized with OU stiffness $\gamma = 1.5$, minimum noise level $\sigma_{\min} = 0.05$, and maximum noise level $\sigma_{\max} = 0.5$. Models are trained using the Adam optimizer ($\text{lr} = 1{\times}10^{-4}$, batch size 8) with Exponential Moving Average (EMA, decay 0.999~\cite{polyak1992acceleration}) across two NVIDIA A100 GPUs for 60 epochs. Inference uses the predictor--corrector reverse SDE sampler with $N{=}30$ discretization steps, $M{=}1$ annealed Langevin corrector step per predictor step, corrector SNR factor $r{=}0.5$, and reverse-time truncation $\epsilon{=}0.03$. The hybrid sampler of \secref{subsec:hybrid_sampling} reuses these identical settings on both sides of the switch.

\subsection{Baseline Methods}
\label{subsec:baselines}

\noindent\textbf{SGMSE (Audio-Only Control).} An audio-only score-based diffusion baseline following the unmodified SGMSE architecture~\cite{sgmse}, sharing the identical NCSN++ backbone, complex STFT representation, training loss, and hyperparameters as MiDiff, but operating strictly on the degraded audio $\mathbf{y}$ without MIDI conditioning (65.6M parameters). Crucially, this model is trained from scratch on the exact same StemGMD drum dataset and under the identical degradation pipeline, serving as the direct counterfactual to evaluate the contribution of symbolic conditioning.

\vspace{2pt}\noindent\textbf{Cold Diffusion.} The recently proposed Cold Diffusion framework for percussive dereverberation~\cite{cold2026percussive} (54.7M parameters), which replaces stochastic noise processes with deterministic degradation models using a $\Delta$-normalized residual UNet architecture, retrained under the identical percussive degradation pipeline.

\subsection{Evaluation Metrics}
\label{subsec:evaluation_metrics}

We evaluate the performance of the proposed algorithm with three complementary metrics; lower is better for all three.

\vspace{2pt}\noindent\textbf{Fr\'echet Audio Distance (FAD)}~\cite{kilgour2018frechet}. 
Our primary perceptual metric is the squared Wasserstein-2 distance between multivariate Gaussian distributions fitted to VGGish~\cite{hershey2017cnn} embeddings of the enhanced and clean reference audio. As a distributional metric, it does not penalize sample-level differences between the enhanced and reference signals, provided their overall acoustic distributions remain similar. This property is particularly important when evaluating generative models that resynthesize transients rather than reconstructing them exactly.

\vspace{2pt}\noindent\textbf{Mel Spectral Distance (MSD)}. Quantifies frame-wise spectral fidelity between the enhanced signal $\hat{\mathbf{x}}$ and the clean reference $\mathbf{x}_0$ in the log-mel domain:
\begin{equation}
\mathrm{MSD}(\mathbf{x}_0, \hat{\mathbf{x}}) = \frac{1}{2\,BK}\sum_{b=1}^{B}\sum_{k=1}^{K} \bigl| L_0[b,k] - \hat{L}[b,k] \bigr|,
\end{equation}
where $L[b,k] = 10\log_{10}\bigl(P[b,k]/\max_{b',k'}P[b',k']\bigr)$ is the power mel spectrogram in dB, referenced to its own maximum, $B{=}80$ mel bands span 0--8\,kHz (1024-point FFT, hop 256), and $K$ is the number of frames. MSD is therefore expressed in dB. The log-magnitude domain aligns with perceptual loudness and reduces the dominance of high-energy bins.

\vspace{2pt}\noindent\textbf{Transient Timing Error (TTER)} measures the mean temporal displacement between drum onsets in the clean reference and their matched counterparts in the enhanced output, normalized by duration (ms/s):
\begin{equation}
\mathrm{TTER}(\mathbf{x}_0, \hat{\mathbf{x}}) = \frac{10^3}{L \cdot |\mathcal{M}|} \sum_{(t_i, \hat{t}_j) \in \mathcal{M}} |t_i - \hat{t}_j|,
\end{equation}
where $L$ is the clip duration in seconds and $\mathcal{M}$ is a one-to-one set of onset pairs. Onsets are detected in both signals with the spectral-flux onset detector of \texttt{librosa} (10\,ms hop, minimum spacing 20\,ms), and $\mathcal{M}$ is built in a single pass over the two time-sorted onset lists: the current reference and enhanced onsets are paired if they lie within $\tau = 50\,\text{ms}$ of each other, and otherwise the earlier of the two is skipped. TTER is computed per clip and averaged over clips; a clip with no matched pair would be excluded, which did not occur for any model. Because the detector is class-agnostic, TTER measures broadband transient timing and does not distinguish drum classes.

We report validation FAD (10 clips, evaluated after each epoch) and test FAD, MSD, and TTER (100 clips, evaluated using the final checkpoints). Checkpoints and hyperparameters are selected exclusively on the validation set. Since FAD estimates are biased for small sample sizes, and selecting the minimum across multiple configurations introduces additional optimism~\cite{gui2024adapting}, the main comparisons are based on the test set. The sole exception is the hybrid switch time, which is swept directly on the test set (\figref{fig:fad_vs_switch}), making its reported optimum optimistic. The classifier guidance and constant-velocity sweeps are evaluated only on the validation set. Absolute FAD values are not directly comparable across splits because validation uses 10 clips and test uses 100. In particular, small-sample FAD exhibits an upward bias that depends on the embedding statistics of each model~\cite{gui2024adapting}. We therefore restrict model comparisons to the same split and avoid interpreting changes in FAD for individual models across splits, such as the increase observed for SGMSE and the decrease for MiDiff from validation to test.

\section{Experimental Results}\label{sec:experimental_results}

\subsection{Main Results}
\label{subsec:main_results}

\begin{figure}[!t]
    \centering
    \includegraphics[width=0.66\linewidth]{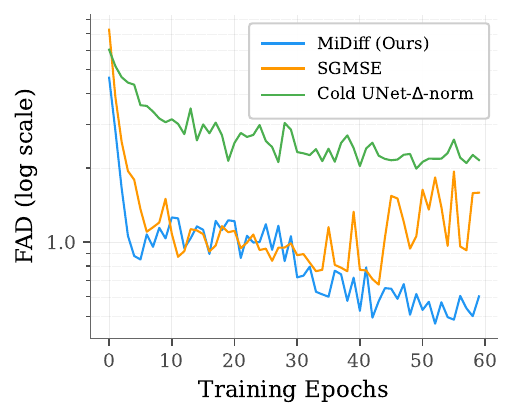}
    \caption{Validation Fr\'echet Audio Distance (FAD, log scale) across training epochs for MiDiff and the SGMSE baseline. Final-checkpoint validation and test FAD are reported in \tabref{tab:main_results}.}
    \label{fig:results}
\end{figure}

\begin{table*}[!t]
\caption{Results on StemGMD. All metrics: lower is better. $^\dagger$On test, Cold Diffusion is run on exactly the same degraded inputs as the other models; its validation inputs are a separately generated 44.1\,kHz version of the 10 degraded clips, resampled to 16\,kHz, so its validation column is only approximately comparable.}
\label{tab:main_results}
\centering
\begin{tabular}{@{}lccccccc@{}}
  \toprule
  & & \multicolumn{2}{c}{\textbf{FAD} $\downarrow$} & \multicolumn{2}{c}{\textbf{MSD (dB)} $\downarrow$} & \multicolumn{2}{c}{\textbf{TTER (ms/s)} $\downarrow$} \\
  \cmidrule(lr){3-4} \cmidrule(lr){5-6} \cmidrule(lr){7-8}
  \textbf{Model} & \textbf{Params} & Val. & Test & Val. & Test & Val. & Test \\
  \midrule
  \textbf{MiDiff (Ours)} & {71.2M} & \textbf{0.467} & \textbf{0.233} & \textbf{1.32} & 1.31 & 0.031 & 0.029 \\
  SGMSE~\cite{sgmse} & 65.6M & 0.673 & 1.268 & 1.39 & \textbf{1.30} & \textbf{0.028} & \textbf{0.028} \\
  Cold UNet-$\Delta$-norm~\cite{cold2026percussive}$^\dagger$ & \textbf{54.7M} & 2.657 & 2.535 & 2.39 & 3.78 & 0.090 & 0.126 \\
  \bottomrule
\end{tabular}
\end{table*}

As shown in \tabref{tab:main_results} and \figref{fig:results}, MiDiff outperforms SGMSE and Cold Diffusion in perceptual quality, reducing FAD relative to SGMSE by 31\% on validation and 82\% on test. The two score-based models achieve comparable spectral and timing performance. MiDiff yields lower validation MSD (1.32 vs.\ 1.39\,dB), while their test values differ by less than 0.01\,dB. SGMSE achieves marginally lower TTER (0.028 vs.\ 0.031\,ms/s on validation and 0.028 vs.\ 0.029\,ms/s on test). For the test clips (41\,s on average), these values correspond to a mean onset displacement of approximately 1.2\,ms, well below the 8\,ms STFT hop. Thus, the benefit of the MIDI prior lies primarily in the acoustic content reconstructed at each onset rather than its timing, as reflected in the FAD improvement. Cold Diffusion (which is the smallest model) exhibits substantially higher timing errors, approximately three times higher on validation (0.090\,ms/s) and four times higher on test (0.126\,ms/s), consistent with its lack of stochastic temporal recovery.

The early-epoch separation in \figref{fig:results} is particularly informative: the gap between MiDiff and the SGMSE baseline widens within the first ten epochs, suggesting that symbolic conditioning reshapes the optimization landscape from the outset rather than merely refining a near-converged audio-only solution. We note that the SGMSE baseline curve exhibits a slight upward drift in later epochs, consistent with mild overfitting in the absence of the structural regularization provided by the MIDI prior. MiDiff adds only 5.6\,M parameters (71.2\,M total vs.\ 65.6\,M baseline), introducing minimal computational overhead. The velocity ablation in \secref{subsec:velocity_ablation}, where corrupted MIDI at the same parameter count falls below the audio-only baseline, indicates that symbolic content, not capacity alone, drives the gain; a capacity-matched control is future work.

Audio-only models, whether stochastic (SGMSE) or deterministic (Cold Diffusion), must infer onsets and instrument identity from the degraded observation alone; MiDiff's piano-roll conditioning supplies this structure, yielding a large FAD gain at comparable MSD and TTER.

\subsection{Qualitative Analysis}

\begin{figure}[!t]
  \centering
  \includegraphics[width=0.78\linewidth]{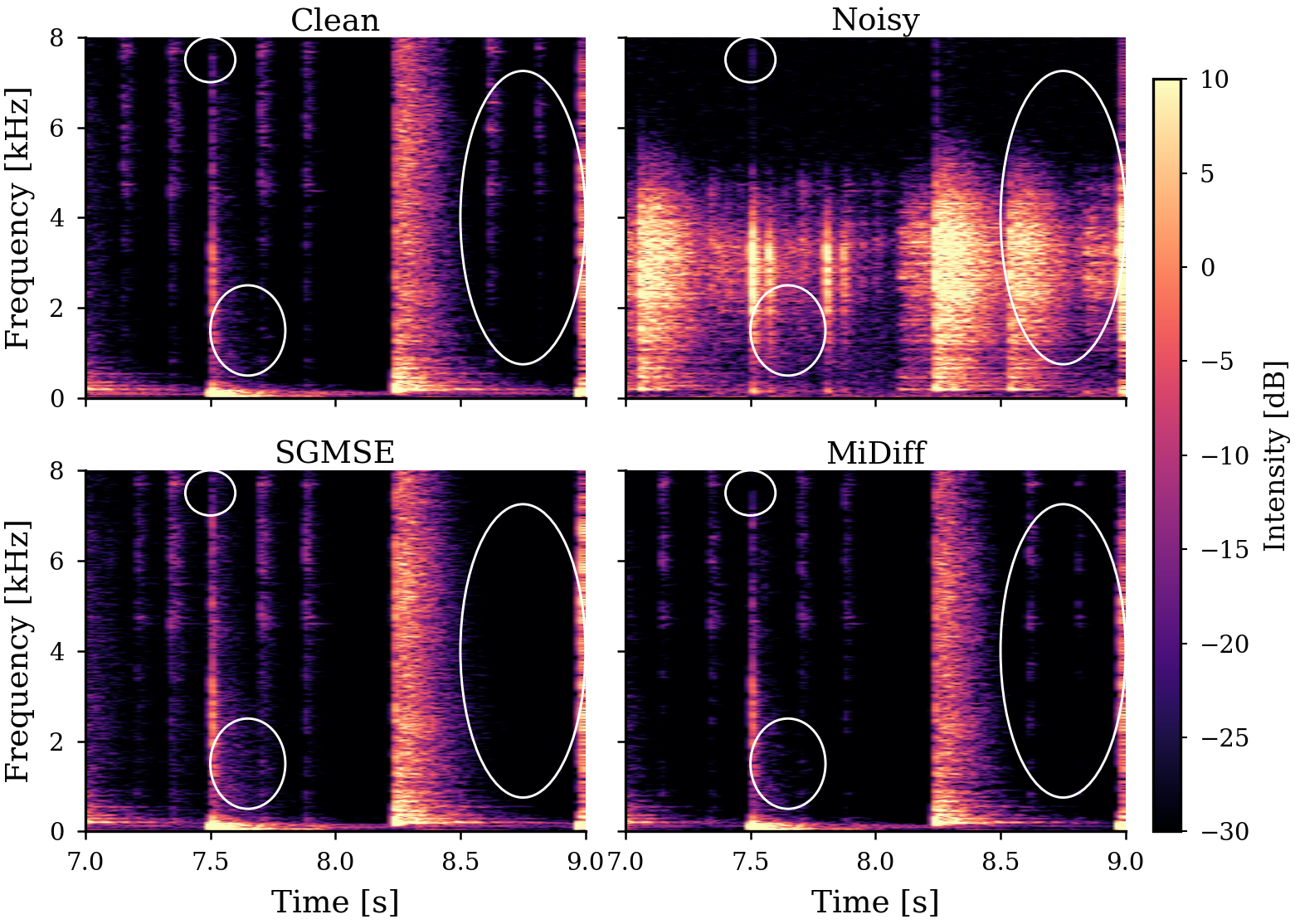}
  \caption{Qualitative STFT spectrogram comparison on a StemGMD test excerpt. Top-left: clean reference; top-right: degraded input; bottom-left: SGMSE baseline; bottom-right: MiDiff. Panels span 2\,s and 0--8\,kHz (the full band at 16\,kHz), so cymbal energy above 8\,kHz is not shown. White ellipses mark regions where MiDiff better preserves 4--8\,kHz cymbal energy and sharp onsets that SGMSE smears or attenuates.}
  \label{fig:spectrogram_comparison}
\end{figure}

The spectrogram comparison in \figref{fig:spectrogram_comparison} shows that SGMSE tends to produce over-smoothed transients, attenuated high-frequency content, and less distinct boundaries between drum hits. In contrast, MiDiff preserves sharper onsets and better separation between individual drum events. Since both models achieve comparable onset timing (\secref{subsec:main_results}), the difference lies in the acoustic content reconstructed at each onset. Guided by MIDI information on instrument identity and strike strength, MiDiff concentrates energy within each hit rather than smearing it across frames. These observations are limited to visual evidence. Quantifying perceptual attributes such as ``punch'' using objective measures (e.g., per-onset crest factor, attack slope, and integrated loudness according to EBU~R128), together with a formal listening study, is left to future work.

\subsection{Classifier Guidance Analysis}
\label{subsec:classifier_guidance_results}

\begin{figure*}[!t]
  \centering
  \includegraphics[width=0.6\textwidth]{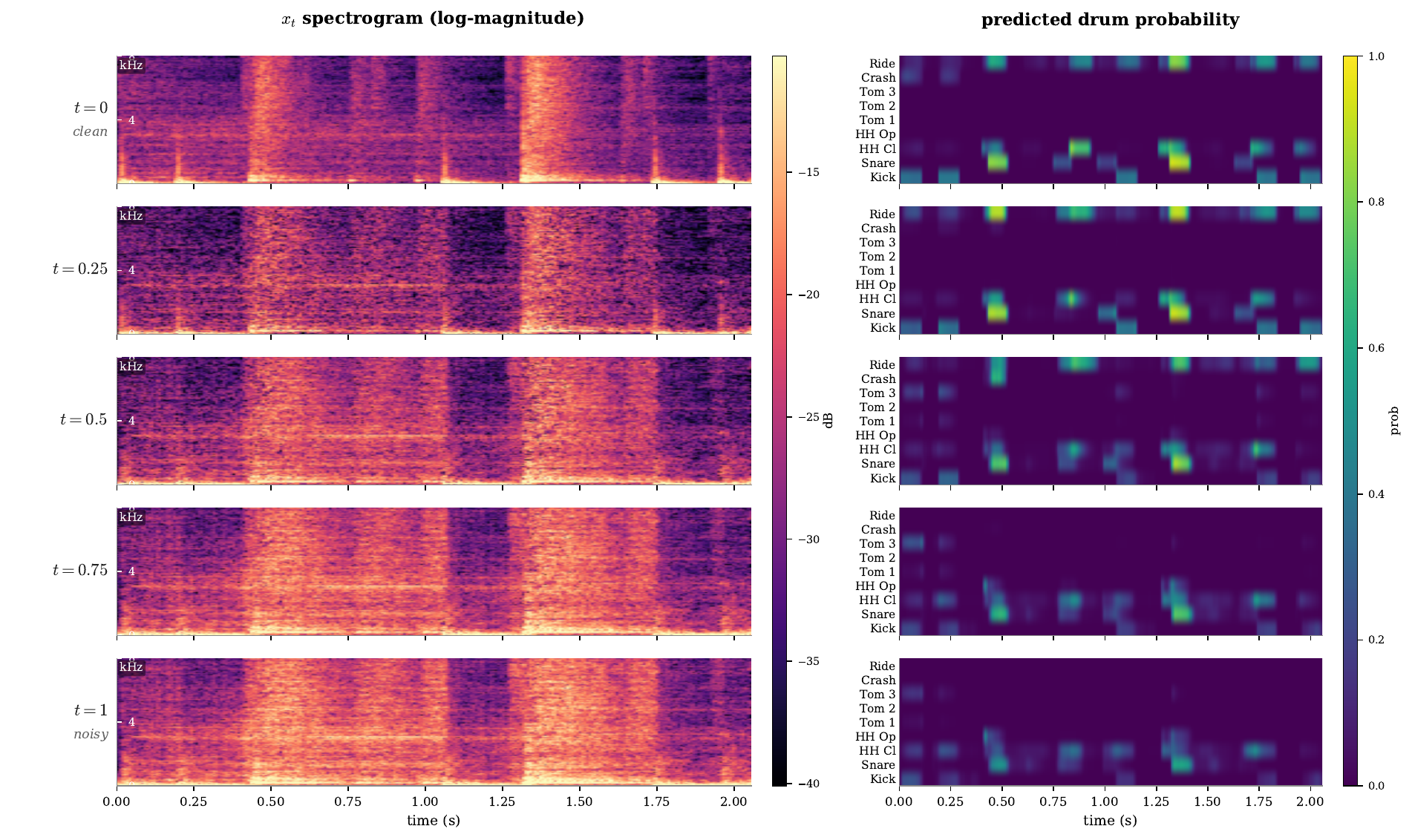}
  \caption{Effect of diffusion time $t$ on the SGMSE forward process and the time-conditional ADT classifier. Each row shows $\mathbf{x}_t$ interpolating between $\mathbf{x}_0$ (top, $t{=}0$) and $\mathbf{y}$ (bottom, $t{=}1$). Right column: per-class onset probabilities $p_\phi(\mathbf{m}\mid\mathbf{x}_t,t)$ remain sharp for small $t$ and drop off sharply as $t \to 1$, since at high noise levels the acoustic observation is too corrupted to carry reliable onset structure. This loss of reliability limits online transcription at large $t$ and motivates the hybrid schedule of \secref{subsec:hybrid_sampling}.}
  \label{fig:adt_progression}
\end{figure*}

\Figureref{fig:adt_progression} illustrates the trained classifier's behavior along the diffusion trajectory. Onset activations remain sharply localized at small $t$ but diminish substantially as the signal approaches $\mathbf{y}$ at $t=1$, as expected, since heavily corrupted observations contain little reliable onset information.

The gradient $\nabla_{\mathbf{x}_t} \log p_\phi(\mathbf{m} \mid \mathbf{x}_t, t)$ guides the reverse process toward spectrograms with higher onset detection probabilities. Since the classifier was trained on signals with well-localized transients, this guidance implicitly promotes sharp, temporally aligned energy concentrations at the onset positions specified by $\mathbf{m}$.

\begin{figure}[!t]
  \centering
  \includegraphics[width=0.64\linewidth]{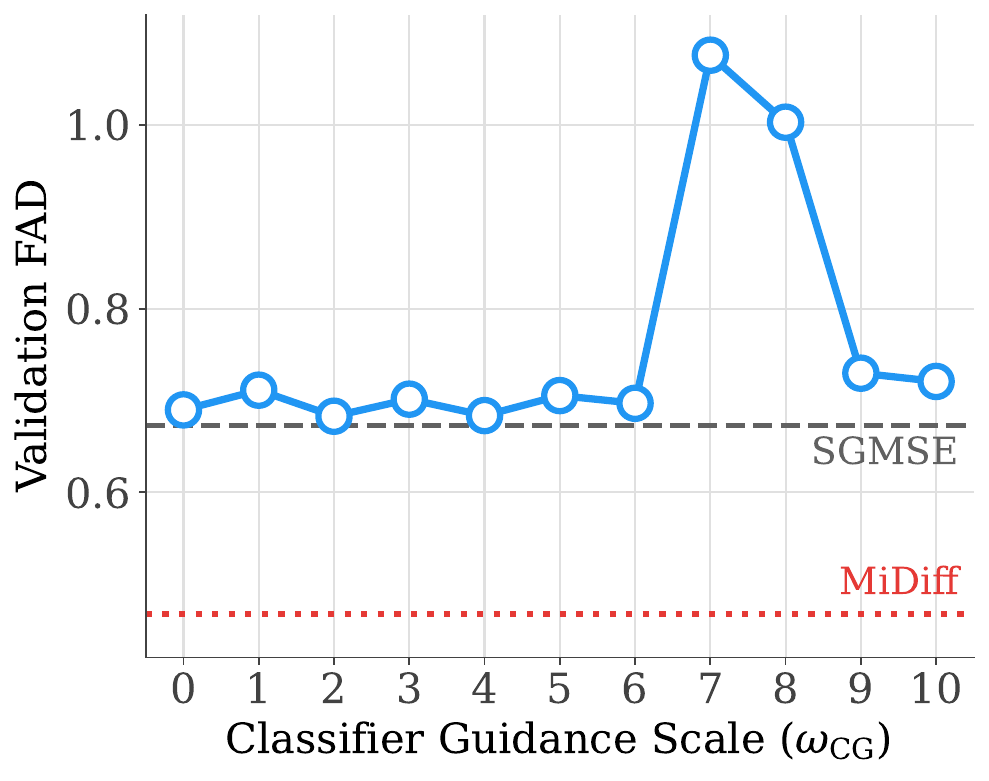}
  \caption{Validation FAD of classifier guidance applied to the audio-only SGMSE score (Eq.~\eqnref{eq:cg_score}) as a function of the scale $w_{\text{CG}} \in \{0,1,\dots,10\}$. The dashed and dotted lines mark SGMSE ($0.673$) and MiDiff conditioned on ground-truth MIDI ($0.467$) from \tabref{tab:main_results}.}
  \label{fig:guidance_cg}
\end{figure}

We evaluate classifier guidance by sweeping $w_{\text{CG}} \in \{0,1,\dots,10\}$, using the time-conditioned ADTOF classifier to guide the audio-only SGMSE score (\figref{fig:guidance_cg}). Without guidance ($w_{\text{CG}}=0$), validation FAD is $0.690$, close to the SGMSE result in \tabref{tab:main_results} ($0.673$), allowing for sampling variation. Classifier guidance provides no meaningful improvement: for $w_{\text{CG}}\leq 6$, FAD ranges from $0.683$ to $0.711$, with the best result ($0.683$ at $w_{\text{CG}}=2$) improving upon the unguided sampler by only $1\%$. Larger scales destabilize the reverse process, increasing FAD to $1.076$ and $1.003$ at $w_{\text{CG}}=7$ and $8$, respectively, before decreasing to $0.72$--$0.73$ at $w_{\text{CG}}\in\{9,10\}$. None approaches MiDiff's FAD of $0.467$, despite both methods using the same ground-truth piano roll. The classifier gradient encourages detectable onsets at the specified positions but provides no direct information about their timbre or energy. These results suggest that symbolic information is more effective when incorporated into the score network than when applied externally during inference. Since classifier guidance provides no meaningful validation improvement, it was not evaluated on the test set.

\subsection{Inference with Transcribed MIDI}
\label{subsec:hybrid_results}

\begin{figure}[!t]
  \centering
  \includegraphics[width=0.64\linewidth]{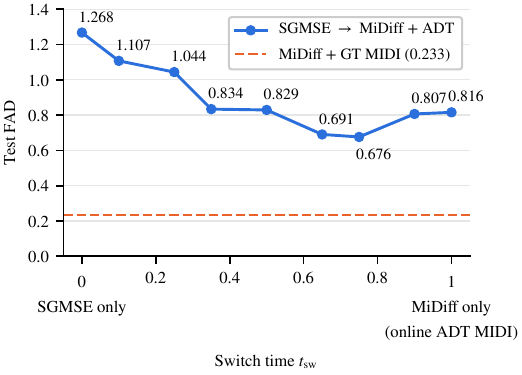}
  \caption{Test FAD of the hybrid SGMSE\,$\rightarrow$\,MiDiff sampler (Algorithm~\ref{alg:hybrid_pc}) as a function of the switch time $t_{\text{sw}}$, with the conditioning piano roll transcribed online from the diffusion state by the time-conditional ADT network. The extremes are the individual models: $t_{\text{sw}}{=}0$ is audio-only SGMSE and $t_{\text{sw}}{=}1$ is MiDiff conditioned on transcribed MIDI at every reverse step. The dashed line marks MiDiff conditioned on ground-truth MIDI ($0.233$, \tabref{tab:main_results}); $t_{\text{sw}}{=}0$ is SGMSE ($1.268$). FAD is minimized at $t_{\text{sw}}{=}0.75$ ($0.676$).}
  \label{fig:fad_vs_switch}
\end{figure}

We evaluate the hybrid sampler of \secref{subsec:hybrid_sampling} by sweeping the switch time over nine values in $[0,1]$, without retraining either score network and using only MIDI generated by the ADT transcriber. The resulting FAD curve (\figref{fig:fad_vs_switch}) is distinctly U-shaped. At $t_{\text{sw}}=0$, the sampler reduces to the audio-only baseline, yielding FAD $1.268$. Increasing the switch time progressively reduces FAD to a minimum of $0.676$ at $t_{\text{sw}}=0.75$, selected on the test set (\secref{subsec:evaluation_metrics}). This represents a $47\%$ improvement over the audio-only baseline without ground-truth MIDI. Beyond this point, FAD increases to $0.807$ at $t_{\text{sw}}=0.9$ and $0.816$ at $t_{\text{sw}}=1$, making the hybrid sampler $17\%$ better than conditioning on transcribed MIDI throughout the trajectory. However, neighboring switch times yield similar results ($0.676$ at $0.75$ vs.\ $0.691$ at $0.65$), and each point represents a single sampling run (\secref{subsec:limitations}).

We attribute this improvement to MiDiff's MIDI conditioning. At $t_{\text{sw}}=0.75$, the state is sufficiently clean for the online transcription to provide useful MIDI information, despite its inaccuracies. Switching at higher noise levels introduces transcription errors that propagate through the reverse process, whereas switching at lower noise levels misses opportunities to exploit informative MIDI conditioning. The remaining gap relative to ground-truth MIDI conditioning ($0.676$ vs.\ $0.233$) suggests further improvements may be achievable with a more accurate transcriber.

\section{Ablation Study}\label{sec:ablation}

\subsection{Conditioning Strategy Ablation}
\label{subsec:conditioning_strategy}

We investigate the three conditioning pathways through two complementary experiments (\tabref{tab:conditioning_ablation}). \emph{Trained alone}: each pathway is trained separately from scratch, using the same encoder, data, and training schedule as MiDiff, and evaluated on validation at its best checkpoint. \emph{Disabled in MiDiff}: the trained MiDiff model is evaluated on the test set with only one pathway active, assessing its individual contribution within the jointly trained model.

Both experiments demonstrate that no single pathway is sufficient. When trained alone, all pathways perform worse than the audio-only baseline on validation (1.39--1.70 vs.\ 0.673) throughout training, from the second epoch onward, whereas their combination achieves 0.467. This indicates that the pathways complement one another during learning. When only one pathway is retained in the trained MiDiff model, test FAD approximately doubles relative to the full model (0.42--0.49 vs.\ 0.233), although it remains substantially below the audio-only baseline (1.268). FiLM retains the greatest benefit (0.419), followed by early injection (0.469) and cross-attention (0.488).

\begin{table}[!t]
\caption{Conditioning Strategy Ablation (FAD, lower is better). Trained alone: a separate model per pathway, validation. Disabled in MiDiff: the full model with the other pathways switched off at inference, test.}
\label{tab:conditioning_ablation}
\centering
\begin{tabular}{@{}lcc@{}}
\toprule
& \textbf{Trained alone} & \textbf{Disabled in MiDiff} \\
\textbf{Pathway kept} & (Val.) & (Test) \\
\midrule
None (SGMSE baseline) & 0.673 & 1.268 \\
Early injection only & 1.695 & 0.469 \\
FiLM only & 1.554 & 0.419 \\
Cross-attention only & 1.394 & 0.488 \\
\textbf{All three (MiDiff)} & \textbf{0.467} & \textbf{0.233} \\
\bottomrule
\end{tabular}
\end{table}

A plausible account is that the mechanisms act at distinct levels: an additive bias at the input (early injection), multiplicative gating throughout the U-Net (FiLM), and content-based retrieval of symbolic frames (cross-attention).

\subsection{Impact of MIDI Velocity Representation}
\label{subsec:velocity_ablation}

\begin{figure}[!t]
  \centering
  \includegraphics[width=0.64\linewidth]{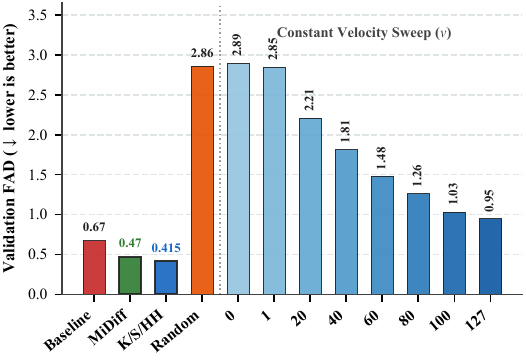}
  \caption{Validation FAD by velocity configuration. Baseline (SGMSE), MiDiff (full kit authentic velocity), K/S/HH (kick/snare/hi-hat authentic velocity), Random velocity, and constant velocities $v \in \{0, 1, 20, 40, 60, 80, 100, 127\}$.}
  \label{fig:velocity_ablation}
\end{figure}

To assess the role of dynamic strike velocity, we compare authentic MIDI velocities with modified configurations (\tabref{tab:midi_configs} and \figref{fig:velocity_ablation}). Restricting conditioning to kick, snare, and hi-hat (K/S/HH) yields validation and test FAD values of 0.415 and 0.216, respectively, slightly outperforming full-kit conditioning (0.467 and 0.233). This may be attributed to the dominance of these instruments, which account for 72.7\% of training onsets and exhibit characteristic high-energy transients. The remaining instruments occur less frequently and often produce softer, more diffuse onsets. Nevertheless, full 9-piece conditioning remains preferable for restoring infrequent fills and cymbal crashes with symbolic guidance. Randomizing note velocities substantially degrades performance (FAD 2.857), as the resulting dynamics conflict with the acoustic energy envelope. With constant velocity, FAD decreases from 2.894 at $v=0$ to 0.946 at $v=127$, suggesting that higher velocities provide stronger onset energy cues. However, even $v=127$ performs worse than the audio-only baseline (0.673), indicating that authentic dynamic MIDI is essential for effective conditioning.

\begin{table}[!t]
\caption{Experimental MIDI Velocity Configurations.}
\label{tab:midi_configs}
\centering
\begin{tabularx}{\linewidth}{@{} >{\raggedright\arraybackslash}p{2.9cm} >{\raggedright\arraybackslash}X @{}}
\toprule
\textbf{Configuration} & \textbf{Description} \\
\midrule
Baseline (Audio-Only) & Standard SGMSE without MIDI conditioning. \\
Authentic Velocity & Full 7-bit MIDI velocity dynamics normalized to $[0,1]$. \\
Kick, Snare, Hi-Hat Only & Velocity active only for kick, snare, and hi-hats. \\
Random Velocity & Uniformly randomized event velocity values. \\
Constant Velocity & Fixed velocity setting ($v \in \{0, 1, 20, 40, 60, 80, 100, 127\}$). \\
\bottomrule
\end{tabularx}
\end{table}

\subsection{Per-Class Performance Breakdown}
\label{subsec:per_class_fad}

\begin{figure}[!t]
  \centering
  \includegraphics[width=0.64\linewidth]{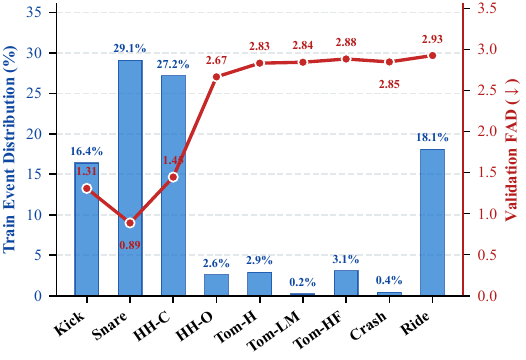}
  \caption{Per-class training event distribution (blue bars, \% of 43,940 note onsets) versus validation FAD (red line, lower is better). HH-C/HH-O: closed/open hi-hat; Tom-H, Tom-LM, Tom-HF: High, Low-Mid and High Floor Tom. Core rhythmic classes (Snare, Kick, Closed Hi-Hat) dominate training occurrences and achieve the lowest FAD scores.}
  \label{fig:per_class_fad}
\end{figure}

To evaluate the contribution of each drum class, we condition the model exclusively on its MIDI events, masking all other classes, and compute the validation FAD of the full output track (\figref{fig:per_class_fad}). FAD generally decreases with the frequency of each class in the training data, except for cymbals with sustained decay. The three dominant rhythmic instruments, Snare, Kick, and Closed Hi-Hat, account for 72.7\% of the 43,940 training onsets and achieve the lowest FAD values (0.886, 1.307, and 1.445, respectively). Snare performs best, benefiting from its 29.1\% share and prominent broadband transients. In contrast, infrequent classes, including High Tom (2.9\%), Low-Mid Tom (0.2\%), High Floor Tom (3.1\%), and Crash (0.4\%), yield substantially higher FAD values (2.83--2.88). Sustained cymbals present an additional challenge, regardless of their frequency of occurrence, as reflected in the FAD values for Ride (2.93) and Open Hi-Hat (2.67). Recovering their metallic decay tails, attenuated by the microphone low-pass filtering (3.2--7.5\,kHz cutoff), is more difficult than restoring the localized transients of kick and snare. This limitation is further constrained by the 8\,kHz Nyquist frequency (\secref{subsec:limitations}).

\section{Discussion}\label{sec:discussion}

\subsection{Why Does MIDI Help?}

Timing is not where MIDI helps: both score-based models place onsets at frame resolution (\secref{subsec:main_results}), and classifier guidance, which only makes the specified onsets detectable, yields no gain (\secref{subsec:classifier_guidance_results}). We therefore attribute the benefit to telling the score network \emph{which} instrument sounds at each onset and \emph{how hard}, so that it reconstructs the corresponding timbre and energy rather than inferring them from the degraded audio. The velocity ablation supports this view: authentic dynamics help, whereas randomized velocities contradict the acoustic energy envelope.

\subsection{Limitations}
\label{subsec:limitations}

\noindent\textbf{Oracle conditioning.} 
StemGMD renders its stems from the conditioning MIDI, making results obtained with ground-truth MIDI an upper bound on performance with MIDI obtained from triggers or transcription, which may contain timing jitter, missed onsets, or spurious events. We examine imperfect MIDI through three experiments: corrupted dynamics (random and constant velocities, \secref{subsec:velocity_ablation}), missing onsets (K/S/HH-only and single-class conditioning, \secref{subsec:velocity_ablation} and \secref{subsec:per_class_fad}), and realistic transcription errors (hybrid sampling, \secref{subsec:hybrid_results}). Timing jitter and spurious onsets are not evaluated separately.

\noindent\textbf{Synthetic data.} 
The clean target is sample-library playback, while degradation is simulated by convolving the summed mixture with a synthetic impulse response comprising decaying noise and a single echo, rather than a realistic room response. This assumes a point-source model for spatially extended instruments. Codec compression is not simulated, and no real recordings are evaluated. Moreover, training and test data come from different sessions of the same drummer and share kit libraries (\secref{subsec:datasets}), potentially yielding optimistic FAD relative to drummer- or kit-disjoint evaluation.

\noindent\textbf{Bandwidth.} 
Two limitations apply: the microphone low-pass ($3.2$--$7.5$\,kHz) removes upper-frequency cymbal content that must be regenerated, while the $16$\,kHz sampling rate imposes an $8$\,kHz Nyquist ceiling on restoration and evaluation. Both affect the cymbal classes in \secref{subsec:per_class_fad}. Full-band restoration, as in~\cite{moliner2023solving}, remains future work.

\noindent\textbf{Single runs.} All results come from one stochastic sampling run, so differences of a few percent are not resolved. 

\noindent\textbf{What MIDI cannot specify.} Room tone, microphone character, performer timbre and within-hit dynamics must still be recovered from the observation.

\section{Conclusions}\label{sec:conclusions}

We presented MiDiff, a MIDI-conditioned score-based diffusion framework for drum audio enhancement. On StemGMD, MiDiff achieves validation and test FAD values of 0.467 and 0.233, outperforming SGMSE (0.673 and 1.268) by 31\% and 82\%, respectively. For scenarios without ground-truth MIDI, we introduced a hybrid reverse schedule that transcribes the diffusion state online and switches from the audio-only score to MiDiff. At $t_{\text{sw}}=0.75$, selected on the test set, it achieves a test FAD of 0.676, a 47\% improvement over the audio-only baseline. These results demonstrate the benefits of MIDI conditioning even when only an audio recording is available.

\section*{AI Usage Acknowledgment}

Generative AI tools were used solely for code debugging (GitHub Copilot and Gemini) and language polishing (ChatGPT). All scientific contributions, including experimental design, model architecture, and result analysis, were made by the authors, who reviewed all AI-generated suggestions and take full responsibility for the manuscript's content.

\bibliographystyle{IEEEtran}
\bibliography{refs}

\end{document}